\documentstyle[12pt,amsfonts,amssymb,latexsym]{article}

           \newcommand{\be}{\begin{equation}}
            \newcommand{\ee}{\end{equation}}
            \newcommand{\bee}[1]{\begin{equation}\label{#1}}
            \newcommand{\bey}{\begin{eqnarray}}
            \newcommand{\byy}[1]{\begin{eqnarray}\label{#1}}
            \newcommand{\eey}{\end{eqnarray}}
            \newcommand{\R}[1]{(\ref{#1})}
            \newcommand{\C}[1]{\cite{#1}}

            \newcommand{\mvec}[1]{\mbox{\boldmath{$#1$}}}

            \newcommand{\ro}{\stackrel{\ \circ}{\varrho}}

            \newcommand{\st}[2]{\stackrel{_#1}{#2}}
            \newcommand{\td}{{^{\bullet}}}
            \newcommand{\dm}{\diamond}
            \newcommand{\D}[1]{\st{\dm}{#1}}
            \newcommand{\trido}{\triangledown}

\begin{document}
\date{\empty}
\title{Phenomenological
Non-Equilibrium \\ Quantum Thermodynamics \\ based on
Modified von Neumann Equations
\thanks{In memory of Robert D. Axelrad}\\
}
\author{W. Muschik\footnote{Corresponding author:
muschik@physik.tu-berlin.de}
\\
Institut f\"ur Theoretische Physik\\
Technische Universit\"at Berlin\\
Hardenbergstr. 36\\D-10623 BERLIN,  Germany}
\maketitle

\noindent
{\bf Abstract:}
The wide-spread opinion is that original quantum mechanics is a reversible theory, but this statement
is only true for undecomposed systems that are those systems which sub-systems are out of
consideration. Ta\-king sub-systems into account, as it is by definition necessary for decomposed
systems, the interaction Hamiltonians which are absent in undecomposed systems generate
irreversibility. Thus, the following two-stage task arises: How to modify von Neumann's equation
for undecomposed systems so that irreversibility appears, and how this modification affects 
decomposed systems\nolinebreak[4] ? The first step was already done in \C{MU19} and
is repeated below, whereas the se\-cond step to formulate a quantum thermodynamics of
decomposed systems is performed here by modifying the von Neumann equations of the
sub-systems by a procedure wich is similar to that of Lindblad's equation \C{LI76}, but different
because the sub-systems interact with one another through partitions.
\vspace{.4cm}\newline

\section*{Preface}

Thermodynamics can be formulated as a field theory or differently as going on in discrete systems
\C{MU04,KES79,PAP20}.
Especially there is the non-equilibrium system itself and its non-equilibrium environment
which are separated by a partition, thus representing a bipartite system with an interior partition.
Consequently, three Hamiltonians appear, each one for the system, the environment and for the
partition which needs beyond its Hamiltonian a thermodynamic specification (called non-inertness)
characterizing, if the partition is heating or cooling. The equation of motion of this bipartite system is
the unchanged Schr\"odinger equation without allowing thermodynamic alterations. Introducing the
corresponding statistical operator, the possibility to establish thermodynamic items emerge:
in original quantum mechanics, the statistical weights of statistical operator are set to be time
independent. This presupposition is cancelled here by introducing a so-called propagator which
consists of the time derivatives of the statistical weights and which changes the composition of the
statistical operator in time. How this propagator allows to introduce thermodynamics is discussed
in the paper. 

\tableofcontents

\section{Introduction}

Referring to the tutorial papers \C{STR1,STR2}, a "theoretical frame work, in which thermodynamic
principles are explained and supplemented by quantum-me\-chanical and statistical considerations,
is called quantum thermodynamics". Beyond that, such a coupling of quantum mechanics and
thermodynamics has to define quantum mechanically the thermodynamical basic quantities such as
power-, heat- and entropy-exchange, entropy and entropy production .

In this paper, quantum mechanics is introduced by two items:
a Hamiltonian which belongs to a Schottky system\footnote{A Schottky system is a discrete system,
that means a "box", which is in interaction with its environment. Many of the devices of process
engineering are Schottky systems.} \C{SCHO29,MUASP,MU18} and by a von Neumann equation
which is modified by a so-called {\em propagator} \C{MU19} which allows to introduce
the distinction between isolated and non-isolated systems\footnote{An isolated (Schottky) system
has no interaction with its environment, a non-isolated one may interact with its environment by
power- heat- and material-exchange.}. The here considered systems belong to two classes: the
system may be {\em undecomposed} \C{MU19}, that means none of its sub-systems is taken into
account, and it may be {\em decomposed}, that means sub-systems are coming into play. Being
decomposed or undecomposed are kinds of description of a Schott\-ky system, thus characterizing
different degrees of knowledge of the system. The interaction between an undecomposed system and
its environment is described {\em semi-classical} \C{MU19}, that means no interaction Hamiltonian is
introduced for the present, a step which is rejected for decomposed systems later on.

Beyond classifying the systems into decomposed and undecomposed ones, there are some other
items which are discussed differently in this paper as usual \C{STR1}:\\
(i) the system's environment is
not necessarily an equilibrium heat reservoir, but may be as the system itself a non-equilibrium system,\\
(ii) not only the Hamiltonian
of the system depends on the protocol (power variables), but also do the Hamiltonians of the
environment and of the interaction,\\
(iii) the Hamiltonian does not belong to a so-called "universum" outside the system, but is
always the Hamiltonian of the system itself,\\
(iv) consequently, there is no "effective non-equilibrium
temperature of the universe" which is defined by the internal energy of the universe (?), but the
non-equilibrium temperature of the system, the {\em contact temperature}
\C{MU77,MUBR77,MU09,MU14}, is used which does not depend on the internal energy,\\
(v) use of the system's contact temperature makes possible to define the {\em entropy exchange}
(entropy flux) independently of an external heat reservoir,\\
(vi) in original quantum mechanics, the heat exchange between two sub-systems vanishes with the
interaction Hamiltonian, thus ignoring active caloric properties of the partition between the
sub-systems. In quantum thermodynamics, a so-called {\em non-inertness} can be introduced by
using the propagator describing whether the partition is cooling or heating,\\
(vii) statistical operators of the sub-systems of a decomposed system are gene\-rated by tracing, thus
introducing the partial Shannon entropy of a sub-system as a state function. In original quantum
mechanics, this Shannon entropy is time independent for isolated systems, a fact which is not true
in quantum thermodynamics because of a {\em modified von Neumann equation}. In contrast to the
observational entropy \C{STR1}, the definition of the Shannon entropy does not depend on an
observer's coarse graining which is not a system property.\\
(viii) The modification of von Neumann's equation which does not change Schr\"odinger's
equation represents the core of the paper. This is achieved by introducing time dependent weights of
the statistical operator, thus generating irreversibility by changing its composition in time.

A Schottky system can be represented as being composed or decomposed. Because we consider
{\em bipartite systems} in the sequel (decomposed into two sub-systems \#1 and \#2),
the density operators and the Hamiltonians are of the following shape:
\bey\nonumber
\mbox{undecomposed:}&\quad&\varrho,\ {\cal H},\hspace{5.15cm} 
\\ \nonumber
\mbox{decomposed:}&\quad&\varrho,\ \varrho^1,\ \varrho^2,\  {\cal H}=
{\cal H}^1+{\cal H}^2+{\cal H}^{12}.\hspace{1.1cm}
\eey
Here, ${\cal H}^{12}$ is the interaction Hamiltonian belonging to the sub-systems which are in
contact to each other by a partition \#(12) which may have individual properties.

\section{Retrospect: Undecomposed Systems\label{US}}

For introducing thermodynamics into quantum mechanics, we start with a
{\em modified von Neumann equation} \C{MU19,KATZ67}\footnote{$\boxplus :=\otimes$ means:
$\boxplus$ is defined by $\otimes$}
\byy{+6}
\partial_t \varrho\ \equiv\ \st{\td}{\varrho}\ 
=\ -\frac{i}{\hbar}\Big[{\cal H},\varrho\Big]+\ro,\qquad
\ro\ :=\ \sum_j \st{\td}{p}_j|\Phi^j ><\Phi^j |,
\\ \label{+3}
\varrho\ :=\ \sum_j p_j |\Phi^j ><\Phi^j |,\quad 0\leq p_j \leq 1,
\quad \sum_j p_j = 1,\quad\mbox{Tr}\varrho\ =\ 1, 
\eey
by use of pure quantum states $\{|\Phi^j >\}$ which are normalized, complete
and orthogonal. The self-adjoint Hamiltonian ${\cal H}(\mvec{a})$ depends on the
work variables $\mvec{a}$. The modification in contrast to the original quantum
theory consists in the demand
\bee{+7}
\vee\ j:\qquad\st{\td}{p}_j\ \neq\ 0\ \longrightarrow\ \ro\ \neq\ \underline{0}.
\ee

The {\em First Law} writes ($E$=energy, $\st{\td}{W}$=power, $\st{\td}{Q}$=heat exchange,
Tr=trace)
\byy{+9}   
E &:=& \mbox{Tr}({\cal H}\varrho)\ \longrightarrow\ 
\st{\td}{E}\ =\ \mbox{Tr}(\st{\td}{\cal H}\varrho) + 
\mbox{Tr}({\cal H}\st{\td}{\varrho}),
\\ \label{+11} 
\st{\td}{W} &:=& \mbox{Tr}(\st{\td}{\cal H}\varrho),\qquad
\st{\td}{Q}\ :=\ \mbox{Tr}({\cal H}\st{\td}{\varrho}),
\\ \label{+13}
\st{\td}{W} &=& \mbox{Tr}\Big(\frac{\partial{\cal H}}{\partial\mvec{a}}
\varrho\Big)\cdot\st{\td}{\mvec{a}}\ =:\ \mathbf{K}\cdot\st{\td}{\mvec{a}},
\eey
and by taking \R{+6}$_2$ into account
\bee{+12}
\st{\td}{Q}\ =\ \mbox{Tr}\Big({\cal H}\ro\Big)\ =\ 
\mbox{Tr}\Big({\cal H}\sum_j\st{\td}{p}_j|\Phi^j ><\Phi^j |\Big).
\ee

Starting with the  {\em Shannon entropy} \C{1KATZ67,NiCh00}, a state function of the
undecomposed system
\bee{+14}
S(\varrho)\ :=\ -k_B \mbox{Tr} (\varrho\ln\varrho),
\ee
the {\em entropy time rate} becomes according to \R{+6}
\bee{+15}
\st{\td}{S}(\varrho)\ =\ 
-k_B \mbox{Tr}\Big(\ro\ln(Z\varrho)\Big),\quad\forall Z \in R^1_+.
\ee
The {\em entropy exchange}\footnote{This entropy exchange refers to the contact temperature of
the system and not to the thermostatic temperature of an external heat reservoir.} is according to \R{+12}$_2$
\byy{+18}
\Xi\ :=\ \frac{\st{\td}{Q}}{\Theta}\ =\
\mbox{Tr}\Big(\frac{{\cal H}}{\Theta}\ro\Big).
\eey
Here $\Theta$ is the non-equilibrium {\em contact temperature} defined by the {\em defining
inequality} \C{MU09,MU14,MU18a}
\bee{+21}
\Big(\frac{1}{\Theta}-\frac{1}{T^\Box}\Big)\st{\td}{Q}\ \geq\ 0,
\ee
or in words
\begin{center}
\parbox[t]{11cm}{
{\sf Definition:} The system's contact temperature $\Theta$ is that thermostatic temperature
$T^\Box$ of the system's equilibrium environment  for which the net heat exchange $\st{\td}{Q}$
between the system and this environment through an inert partition\footnotemark\ vanishes by
change of sign.}
\end{center}
\footnotetext{"Inert" means: the partition does not emit or absorb power or/and heat.}

The {\em entropy production}\footnote{The entropy of discrete systems satisfies the balance equation \R{+20}$_1$.} becomes according to \R{+15}, \R{+18} and the 2nd law
\bee{+20}
\Sigma\ =\ \st{\td}{S} - \Xi\ =\ 
-\mbox{Tr}\Big\{\Big(\frac{\cal H}{\Theta}+k_B\ln(Z\varrho)\Big)\ro\Big\}
\ \geq\  0.
\ee

If a discrete non-equilibrium system ${\cal G}$ is isolated, that means the partition 
$\partial{\cal G}$ between ${\cal G}$ and its environment ${\cal G}^\Box$ becomes impervious to
heat and power (and matter), the corresponding exchanges are suppressed
\bee{H1}
\st{\td}{W}_{iso}\ \equiv\ 0,\qquad\st{\td}{Q}_{iso}\ \equiv\ 0,
\ee
and the non-exchange quantities of the system as the energy are uneffected by the system's
isolation. That means according to \R{+9}$_1$, the density operator $\varrho$ and the Hamiltonian
$\cal H$ are uneffected by the isolation of the system. Consequently, also the $\{p_j\}$ and the
$\{|\Phi^j>\}$ are according to \R{+3}$_1$ uneffected by the system's isolation.

From \R{H1}$_2$ and \R{+18}$_1$ follows that in {\em isolated systems} the entropy exchange
vanishes
\bee{20a}
\Xi_{iso}\ =\ 0\ =\ \mbox{Tr}\Big(\frac{{\cal H}}{\Theta}\ro_{iso}\Big),
\ee
that means according to \R{+18}, that the act of isolating the system transforms the 
{\em propagator} $\ro$ by changing the time rates of the weights of the density operator \C{MU20}
\bee{+47}
\ro\quad\longrightarrow\quad\ro_{iso}.
\ee
Establishing the definition
\bee{+48}
\ro_{ex}\ :=\ \ro -\ro_{iso},
\ee
we obtain the propagator in
\byy{+49}
\mbox{non-isolated closed systems:}&\qquad&
\ro\ = \ \ro_{ex} + \ro_{iso}
\\ \label{+50}
\mbox{isolated systems:}&\qquad&
\ro_{iso}.
\eey
Isolation of the system causes the split of the propagator $\ro$ \R{+49} into an
{\em exchange part} $\ro_{ex}$ and into a {\em dissipative thermal part}  $\ro_{iso}$.

Because the entropy production is defined as the time rate of entropy in isolated systems
according to \R{+20}$_1$, \R{20a}$_1$ and \R{+15}
\bee{?1}
\Sigma\ :=\ \st{\td}{S}_{iso}\ =\ -k_B \mbox{Tr}\Big(\ro_{iso}\ln(Z\varrho)\Big).
\ee
Inserting the quantum-theoretical relation \R{+20}, \R{?1} results in two expressions for the
entropy production 
\bee{b?1}
\Sigma\ =\ -\mbox{Tr}\Big\{\Big(\frac{\cal H}{\Theta}+k_B\ln(Z\varrho)\Big)\ro\Big\} =
-k_B \mbox{Tr}\Big(\ro_{iso}\ln(Z\varrho)\Big)\ \geq\ 0.
\ee
The first term represents the entropy production in undecomposed non-isolated closed systems
according to \R{+49}, whereas according to \R{+50}, the second term is the entropy production of
the same undecomposed, but isolated system: isolation does not influence the entropy production.
As the entropy \R{+14} and the entropy time rate \R{+15}, also the entropy production
\R{b?1}$_2$ does not depend on the Hamiltonian.

Inserting \R{+49} into \R{b?1} and taking \R{20a} into account results in
\bee{c?1}
0\ =\ -\mbox{Tr}\Big\{\Big(\frac{\cal H}{\Theta}+k_B\ln(Z\varrho)\Big)\ro_{ex}\Big\}\ =\
\st{\td}{S}_{ex}-\Xi_{ex},
\ee
by taking \R{+15} and \R{+18} into account,
an expression which allows to represent the contat temperature of undecomposed systems
quantum theoretically \C{MU20}
\bee{c?1a}
\frac{1}{\Theta}\ =\ \frac{-\mbox{Tr}\Big\{k_B\ln(Z\varrho)\ro_{ex}\Big\}}
{\mbox{Tr}\{{\cal H}\ro_{ex}\}},\qquad\ro_{ex}\neq\underline{0}.
\ee
depicting that contact temperature can only be measured, if the system is not isolated.

Now equilibrium is discussed which is defined by {\em necessary equilibrium conditions}.
These are\footnote{Thermodynamical demands are put between $\blacksquare$ and
$\blacksquare$.}
\byy{d?1}\blacksquare\hspace{1cm}
\mbox{according to \R{+6}:}&&\qquad\st{\td}{\varrho}{^{eq}}\ =\ \underline{0},
\quad\ro{^{eq}}\ =\ \underline{0}, \quad\ro{^{eq}_{iso}}\ =\ \underline{0},\hspace{.4cm}
\\ \label{d?2}
\mbox{according to \R{+13}:}&&\qquad\st{\td}{\mvec{a}}\ =\ \mvec{0}.\hspace{5cm}
\blacksquare
\eey
From these necessary equilibrium conditions follow {\em complementary} ones. These are
\byy{d?3}
\mbox{according to \R{+6}$_1$:}&& [{\cal H},\varrho{^{eq}}]\ =\ \underline{0},
\\ \label{d?4}
\mbox{according to \R{+48}:}&&\ro{^{eq}_{ex}}\ =\ \underline{0}, 
\\ \label{d?5}
\mbox{according to \R{+13}, \R{+12} and \R{+15}:}&&
\st{\td}{W}{^{eq}}=0,\ \st{\td}{Q}{^{eq}}=0,\ \st{\td}{S}{^{eq}}=0,
\\ \label{d?6}
\mbox{according to \R{+18} and \R{+20}$_1$:}&& \Xi{^{eq}}=0,\ \Sigma{^{eq}}=0.
\eey

Now {\em reversible "processes"}, $\Sigma{^{rev}}\equiv\Sigma^{eq} = 0$, are considered according to \R{+6}$_1$
\bee{f?1b2}
\st{\td}{\varrho}{^{rev}}(\tau)
-\ro{^{rev}}(\tau)\ =\ -\frac{i}{\hbar}[{\cal H},\varrho{^{rev}}(\tau)],
\quad\tau\geq 0,
\ee
with the time replacing path parameter $\tau$ along the reversible trajectory and with the initial
conditions\footnote{\R{f?1b3}$_1$ is only valid at $\tau=0$, but is not satisfied for the time
derivatives \R{f?1b3}$_2$ for which \R{f?1b4} is valid.} 
\bee{f?1b3}
\varrho{^{rev}}(0)\ \doteq\ \varrho^{eq}\quad\wedge\quad\st{\td}{\varrho}{^{rev}}(0)\ \neq\ 
\st{\td}{\varrho}{^{eq}}\ =\ 0.
\ee
The initial condition of the time derivative $\st{\td}{\varrho}{^{rev}}(0)$ follows from the
equation of "motion" of the reversible "process" \R{f?1b2}
\bee{f?1b4}
\st{\td}{\varrho}{^{rev}}(0)
-\ro{^{rev}}(0)\ =\ -\frac{i}{\hbar}[{\cal H},\varrho{^{eq}}]\ =\ {\underline 0}, 
\ee
according to \R{f?1b3}$_1$ and \R{d?3}.
From \R{b?1} follows for
\byy{f?1a}
\mbox{non-isolated closed systems:}&&
\mbox{Tr}\Big\{\Big(\frac{\cal H}{\Theta{^{rev}}}+k_B\ln(Z\varrho{^{rev}})\Big)\ro{^{rev}}\Big\}=0,
\hspace{.4cm}
\\ \label{f?1b}
\mbox{isolated systems:}&& \mbox{Tr}\Big(\ro{_{iso}^{rev}}\ln(Z\varrho{^{rev}})\Big)=0.
\eey
Considering \R{f?1a} and \R{f?1b} at $\tau=0$ and taking the initial condition \R{f?1b3}$_1$ into
account results in
\byy{f?1c}
\mbox{non-isolated closed systems:}&&
\mbox{Tr}\Big\{\Big(\frac{\cal H}{\Theta_{eq}}+k_B\ln(Z\varrho^{eq})\Big)\ro{^{rev}}(0)\Big\}=0,
\hspace{.6cm}
\\ \label{f?1d}
\mbox{isolated systems:}&& \mbox{Tr}\Big(\ro{_{iso}^{rev}(0)}\ln(Z\varrho^{eq})\Big)=0.
\eey
Because the initial condition \R{f?1b4} of $\st{\td}{\varrho}{^{rev}}(0)$ can be chosen arbitrarily,
from \R{f?1c} and \R{f?1d} follow necessary, sufficient and excluding each other complementary
equilibrium conditions
\bee{g?1}
\frac{\cal H}{\Theta_{eq}}+k_B\ln(Z\varrho^{eq})\ =\ {\underline 0}
\quad\mbox{or}\quad \ln(Z\varrho^{eq})\ =\ {\underline 0},
\ee
resulting in {\em equilibrium distributions} \C{MU19} for
\byy{52c}
\mbox{isolated systems:}\quad \varrho^{eq}\ =\ \frac{1}{N}\sum_{j=1}^N |\Phi^j_{eq}><{\Phi}^j_{eq}|\ =:\ \varrho_{mic},\quad N<\infty,
\\ \nonumber
\mbox{non-isolated closed systems:}\hspace{7cm}
\\ \label{52d}
\varrho^{eq}\ =\ \frac{1}{Z} 
\exp\Big[-\frac{{\cal H}}{k_B \Theta}_{eq}\Big]\ =:\ \varrho_{can},\quad
Z\ =\ \mbox{Tr}\exp\Big[-\frac{{\cal H}}{k_B \Theta}_{eq}\Big]
\eey
which satisfy \R{d?3}.
Thus, the canonical and micro-canonical distributions are derived without statistical arguments
modifying the von Neumann equation by a thermodynamical induced propagator $\ro$ and initial
conditions of a special reversible process.

Here finishes the brief repetition of the main results in undecomposed systems
characterized by a single density operator and the corresponding Hamiltonian.
Decomposed systems are discussed in the sequel.

\section{Decomposed Bipartite Systems\label{DS}}

\subsection{The density operators}

We consider a Schottky system which is decomposed into two sub-systems, $\#1$ and
$\#2$ \C{MU20}. Each sub-system is described by a partial density operator,
$\varrho^1\ \mbox{and}\ \varrho^2$, and by a partial Hamiltonian, ${\cal H}^1$ and ${\cal H}^2$.
The interaction between the sub-systems is represented by the interaction Hamiltonian
${\cal H}^{12}$ and by a {\em partition} $\#(12)$ between them. The Hamiltonian $\cal H$ of the
undecomposed system \R{+3}$_1$ is the sum of the partial Hamiltonians of the two
sub-systems, ${\cal H}^1$ and ${\cal H}^2$, and of the interaction Hamiltonian
${\cal H}^{12}$
\bee{§1}
(\#1\cup\#2\cup\#(12)):\qquad {\cal H}\ =\ {\cal H}^1+{\cal H}^2+{\cal H}^{12}.
\ee

By choosing 
an orthogonal basis $\{|\Psi^k_1>\}$ belonging to sub-system \#1 and an other one 
$\{|\Psi^l_2>\}$ belonging to the other sub-system \#2,
the tensor product of these bases form an orthogonal basis of the decomposed
system
\bee{+27}
\{|\Psi^k_1>\otimes
|\Psi^l_2>\}\ \equiv\ \{|\Psi^k_1>|\Psi^l_2>\}.
\ee
With respect to the tensorial base \R{+27}, we used  in \R{§1} the abbreviations
\bee{+38}
{\cal H}^1 \otimes I^2\ \equiv\ {\cal H}^1,
\qquad
I^1\otimes{\cal H}^2\ \equiv\ {\cal H}^2,
\quad\longrightarrow\quad
[{\cal H}^1,{\cal H}^2]\ =\ 0,
\ee
with the unity operators $I^i,\ i=1,2,$ belonging to the corresponding factors of 
the tensor product \R{+27}.

The interaction of the sub-systems with the environment
is for the present described {\em semi-classically} \C{MU19}, that means, by power- and
heat-exchanges which result from the partial Hamiltonians of the sub-systems, ${\cal H}^1$ and
${\cal H}^2$, similar as \R{+11}. There is for the present no Hamiltonian of the environment
and also no interaction Hamiltonian describing the interaction between the environment and the
bipartite system. The interaction Hamiltonian ${\cal H}^{12}$ refers exclusively to
the interaction between the two sub-systems and is independent of the system's
environment\footnote{We will get rid of the semi-classical description in sect.\ref{RCD}}.

According to the basis of the bipartite system \R{+27}, the corresponding density
operator is analogous to \R{+3}$_1$
\bee{38a}
\varrho\ =\ \sum_{kl}p_{kl}|\Psi_1^k>|\Psi_2^l><\Psi_2^l|<\Psi_1^k|,\quad
\mbox{Tr}\varrho\ =\ 1,
\ee
and the propabilities $p_{kl}$ satisfy \R{+3}$_{2,3}$ analogously. The propagator
is ana\-logous to \R{+6}$_2$
\bee{38b}
\ro\ =\ \sum_{kl}\st{\td}{p}_{kl}|\Psi_1^k>|\Psi_2^l><\Psi_2^l|<\Psi_1^k|,\quad
\mbox{Tr}\ro\ =\ 0.
\ee

The partial density operators of the sub-systems follow from the density operator
$\varrho$ by tracing
\byy{§2}
\varrho^1 &:=& \mbox{Tr}^2\varrho\ =\ \sum_k\sum_jp_{kj}|\Psi_1^k>
<\Psi_1^k|,
\\ \label{§2a}
\varrho^2 &:=& \mbox{Tr}^1\varrho\ =\ \sum_l\sum_jp_{jl}|\Psi_2^l>
<\Psi_2^l|,
\eey
Analogous definitions as \R{38b} are valid for the propagators $\ro\!{^1}$ and $\ro\!{^2}$
of the sub-systems.

\subsection{Modified von Neumann equations\label{MVNE}}

Using the decomposition of the Hamiltonian \R{§1}$_2$, the modified von Neumann equation
\R{+6}$_1$ of the undecomposed system
\bee{+43}
\st{\td}{\varrho}\ =\ 
-\frac{i}{\hbar} \Big[({\cal H}^1+{\cal H}^2+{\cal H}^{12}),\varrho\Big]+\ro.
\ee
results in two equations of motion for the traced density operators of the sub-systems
by taking \R{59s}$_1$ into account
\bey\nonumber
\st{\td}{\varrho}{^1}\ :=\ \mbox{Tr}^2\st{\td}{\varrho} &=& 
-\frac{i}{\hbar}\mbox{Tr}^2 \Big[{\cal H}^1,\varrho\Big]
-\frac{i}{\hbar}\mbox{Tr}^2 \Big[{\cal H}^{12},\varrho\Big]
+\mbox{Tr}{^2\!\ro}\ =
\\ \label{+44} 
&=& -\frac{i}{\hbar}\Big[{\cal H}^1,\varrho^1\Big]
-\frac{i}{\hbar}\mbox{Tr}^2 \Big[{\cal H}^{12},\varrho\Big]+\ro{^1},
\\  \nonumber
\st{\td}{\varrho}{^2}\ :=\ \mbox{Tr}^1\st{\td}{\varrho} &=& 
-\frac{i}{\hbar}\mbox{Tr}^1 \Big[{\cal H}^2,\varrho\Big]
-\frac{i}{\hbar}\mbox{Tr}^1 \Big[{\cal H}^{12},\varrho\Big]
+\mbox{Tr}^1\ro\ =
\\ \label{+45}
&=&-\frac{i}{\hbar} \Big[{\cal H}^2,\varrho^2\Big]
-\frac{i}{\hbar}\mbox{Tr}^1 \Big[{\cal H}^{12},\varrho\Big]
+\ro{^2}.
\eey

As \R{+44} and \R{+45} depict, the traced operators $\st{\td}{\varrho}\!\!{^1}$
and $\st{\td}{\varrho}\!\!{^2}$ include as expected the interaction Hamiltonian
${\cal H}^{12}$, but not the Hamiltonian of the neighboring sub-system.
Beyond that, also the density operator $\varrho$ \R{+3}$_1$ of the undecomposed system
appears in the rate equations belonging to the sub-systems. That means: first of all, the
equation of motion \R{+43} of the undecomposed system has to be solved.

Introducing a modified propagator
\byy{+45-1}
\st{\trido}{\varrho}\ :=\ -\frac{i}{\hbar}\Big[{\cal H}^{12},\varrho\Big]+\ro,
\hspace{3.5cm}
\\ \label{+45a}
\st{\trido}{\varrho}\!{^1}\ =\ -\frac{i}{\hbar}\mbox{Tr}^2 \Big[{\cal H}^{12},\varrho\Big]+\ro{^1},\quad
\st{\trido}{\varrho}\!{^2}\ =\ -\frac{i}{\hbar}\mbox{Tr}^1 \Big[{\cal H}^{12},\varrho\Big]+\ro{^2},
\eey
the equations of motion \R{+44} and \R{+45} result in
\bee{+45b}
\st{\td}{\varrho}{^1}\ =\  -\frac{i}{\hbar}\Big[{\cal H}^1,\varrho^1\Big]
+\st{\trido}{\varrho}\!{^1},\quad
\st{\td}{\varrho}{^2}\ =\  -\frac{i}{\hbar}\Big[{\cal H}^2,\varrho^2\Big]
+\st{\trido}{\varrho}\!{^2}.
\ee
The shape of the equations of motion \R{+45b} of the sub-sytems is identical
with that of the undecomposed system \R{+6} except for $\ro$ is replaced by
$\st{\trido}{\varrho}\!{^1}$ and $\st{\trido}{\varrho}\!{^2}$. The vanishing of the propagator
$\ro$ in original quantum mechanics does not cause vanishing of the modified propagators
$\st{\trido}{\varrho}\!{^1}$ and $\st{\trido}{\varrho}\!{^2}$. This fact may be a motivation to
consider the modified von Neumann equations \R{+45b} also for decomposed systems of
original quantum mechanics.

In original quantum mechanics ($\ro \equiv 0$) the two sub-systems of the bipartite
system are isolated from each other by $[{\cal H}^{12},\varrho]\equiv 0$ according
to \R{+44} and \R{+45}.
Using the modified von Neumann equation \R{+43}, isolation of the sub-systems is described by
$\st{\trido}{\varrho}\equiv 0$ according to \R{+45b}, resulting in
$\ro = \frac{i}{\hbar}[{\cal H}^{12},\varrho]$: in the case of mutual isolation of the
sub-systems, the propagator $\ro$ removes the quantum mechanical interaction
(${\cal H}^{12}\equiv\hspace{-.4cm}/\hspace{.3cm}0$).

In equilibrium, \R{+44} and \R{+45} yield by taking \R{d?1} into account
\bee{+45c}
0\ =\ -\frac{i}{\hbar}[{\cal H}{^A_{eq}},\varrho{^A_{eq}}]
-\frac{i}{\hbar}\mbox{Tr}^B[{\cal H}{^{12}_{eq}},\varrho{^{eq}}],\qquad A=1,2; B\neq A.
\ee
Taking \R{59s}$_3$ into account, tracing results in
\bee{+45d}
0\ =\ \mbox{Tr}^A[{\cal H}{^A_{eq}},\varrho{^A_{eq}}]
+\mbox{Tr}^A\mbox{Tr}^B[{\cal H}{^{12}_{eq}},\varrho{^{eq}}]\ =\
\mbox{Tr}[{\cal H}{^{12}_{eq}},\varrho{^{eq}}],
\ee
 an expression which replace \R{d?3} in case of decomposed systems. Taking \R{+45d}$_2$
into account, \R{+45-1} and \R{+45b} result in
\bee{+45e}
\st{\trido}{\varrho}{^{eq}}\ =\ \underline{0},\qquad [{\cal H}{^A_{eq}},\varrho{^A_{eq}}]\ =\
\underline{0}.
\ee

Obviously, the propagators which modify the von Neumann equations cannot be established by pure
quantum theoretical arguments, because their introduction should connect thermodynamics and
quantum mechanics. Consequently, we need thermodynamical argumentation to lay down the
pro\-perties of the propagators. This will be done in the next sections in several steps.

\subsection{The exchanges\label{EXCH}}
\subsubsection{Power exchanges\label{PEX}}

The work variables $\mvec{a}$ in \R{+13} belong to the undecomposed system.
Switching over to a decomposed system, these work variables have to be replaced by those
which belong to the sub-systems and their environment, $\mvec{a}^1$ and $\mvec{a}^2$, and
those $\mvec{a}^{12}$ which are related to the interaction between the sub-systems
\bee{38d}
\mvec{a}\ \longrightarrow\ (\mvec{a}^1, \mvec{a}^2, \mvec{a}^{12})
\longrightarrow\ 
{\cal H}(\mvec{a}^1, \mvec{a}^2, \mvec{a}^{12}).
\ee
According to their definition, the work variables of the decomposed system \R{38d}$_2$
are attached to the partial Hamiltonians as follows\footnote{$\mvec{a}^{12}$
describes the position of a partition between the sub-systems which is displaceable thus
influencing the three Hamiltonians.}
\bee{38d1}
{\cal H}^1(\mvec{a}^1,\mvec{a}^{12}),\quad {\cal H}^2(\mvec{a}^2,\mvec{a}^{12}),\quad {\cal H}^{12}(\mvec{a}^{12}),
\ee
resulting in
\byy{38d2}
\st{\td}{\cal H}\!{^1}\ =\ \frac{\partial{\cal H}^1}{\partial\mvec{a}^1}
\cdot\st{\td}{\mvec{a}}\!{^1}
+\frac{\partial{\cal H}^1}{\partial\mvec{a}^{12}}
\cdot\st{\td}{\mvec{a}}\!{^{12}},
&&
\st{\td}{\cal H}\!{^2}\ =\ \frac{\partial{\cal H}^2}{\partial\mvec{a}^2}\cdot\st{\td}{\mvec{a}}\!{^2}
+\frac{\partial{\cal H}^2}{\partial\mvec{a}^{12}}
\cdot\st{\td}{\mvec{a}}\!{^{12}},
\\ \label{38d3}
\st{\td}{\cal H}\!{^{12}} &=& \frac{\partial{\cal H}^{12}}{\partial\mvec{a}^{12}}
\cdot\st{\td}{\mvec{a}}\!{^{12}}.
\eey

Starting with \R{+13}$_1$, we obtain with \R{38d2} and \R{38d3} and the suitable
tracing
\byy{§1z}
\st{\td}{W}&=&
\mbox{Tr}^2\mbox{Tr}^1\Big\{(\st{\td}{{\cal H}}\!{^1}+\st{\td}{{\cal H}}\!{^2}+
\st{\td}{{\cal H}}\!{^{12}})\varrho\Big\}\ =
\\ \label{§2z}
&=&\mbox{Tr}^1(\st{\td}{{\cal H}}\!{^1}\varrho^1)+
\mbox{Tr}^2(\st{\td}{{\cal H}}\!{^2}\varrho^2)+
\mbox{Tr}(\st{\td}{{\cal H}}\!{^{12}}\varrho).
\eey
This decomposition of the time derivative of the Hamiltonian
allows to define external and internal power exchanges: external between each sub-system
and the environment and internal exchanges between the sub-systems themselves
\byy{38d4}
\st{\td}{W}\!{{^A}\!\!_{ex}} &:=&
\mbox{Tr}^A\Big(\frac{{\partial\cal H}^A}{\partial\mvec{a}^A}\varrho^A\Big)
\cdot\st{\td}{\mvec{a}}\!{^A},\quad A=1,2,\quad
\frac{{\partial\cal H}^{12}}{\partial\mvec{a}^A}\ \equiv\ \mvec{0},
\\ \label{38d5}
\st{\td}{W}\!{{^A}\!\!_{int}} &:=&
\mbox{Tr}^A\Big(\frac{{\partial\cal H}^A}{\partial\mvec{a}^{12}}\varrho^A\Big)
\cdot\st{\td}{\mvec{a}}\!{^{12}},\quad\st{\td}{W}\!{{^{12}_{int}}}\ :=\ 
\mbox{Tr}\Big(\frac{{\partial\cal H}^{12}}{\partial\mvec{a}^{12}}\varrho\Big)
\cdot\st{\td}{\mvec{a}}\!{^{12}}.
\eey
Consequently, we obtain according to \R{§2z}
\bee{38d6}
\st{\td}{W}_{ex}\ :=\ \st{\td}{W}\!{{^1}\!\!_{ex}}+\st{\td}{W}\!{{^2}\!\!_{ex}},
\qquad
\st{\td}{W}_{int}\ :=\ \st{\td}{W}\!{{^1}\!\!_{int}}+\st{\td}{W}\!{{^2}\!\!_{int}}+
\st{\td}{W}\!{{^{12}_{int}}}.
\ee

Accepting that the sum of the internal power exchanges is zero, we have
\bee{38e}\blacksquare\hspace{3cm}
\st{\td}{W}_{int}\ =\ \mbox{Tr}\Big(\frac{\partial\cal H}{\partial\mvec{a}^{12}}\varrho\Big)
\cdot\st{\td}{\mvec{a}}\!{^{12}}\equiv\ 0,\hspace{3cm}\blacksquare
\ee
and we obtain from \R{38d6}$_2$
\bee{38e1}
-\st{\td}{W}\!{^1_{int}}\ =\ \st{\td}{W}\!{^2_{int}}+
\st{\td}{W}\!{^{12}_{int}}.
\ee
If $\st{\td}{W}\!{{^{12}}\!\!_{int}}>0$, the partition between the sub-systems
is power absorbing, and if $\st{\td}{W}\!{{^{12}}\!\!_{int}}<0$, it is power supplying.

\subsubsection{Heat exchanges\label{HEX}}

There are three different kinds of heat excchanges: that of the undecomposed system \R{+12}$_1$,
the external heat exchanges taking place between each sub-system and the equilibrium
environment, the internal heat exchanges taking place between the sub-systems themselves.
For the present, the quantum theoretical expressions of these heat exchanges are specified
and their connection to the corresponding contact temperatures is given below.

According to \R{+11}$_2$ and \R{+45b}, the heat exchange
of the sub-system \#A is defined by taking \R{3.3}, \R{+38}$_3$ and \R{+45-1} into account
\bey\nonumber
\st{\td}{Q}{^A} &:=& 
\mbox{Tr}^A({\cal H}^A\st{\td}{\varrho}{^A})\ =\ 
\mbox{Tr}({\cal H}^A\st{\td}{\varrho})\ =\ 
\mbox{Tr}\Big\{{\cal H}^A \Big(\ro
-\frac{i}{\hbar} \Big[{\cal H},\varrho\Big]\Big)\Big\} =\hspace{.3cm}
\\ \nonumber
&=&
\mbox{Tr}\Big\{{\cal H}^A \Big(\ro
-\frac{i}{\hbar} \Big[{\cal H}^{12},\varrho\Big]\Big)\Big\}\ =\
\mbox{Tr}\Big\{{\cal H}^A \st{\trido}{\varrho}\Big\}\ =\
\\ \label{AL3a}
&=&
\mbox{Tr}^A({\cal H}^A\ro\!{^A})
+\frac{i}{\hbar}\mbox{Tr}\Big\{{\cal H}^{12}\Big[{\cal H}^A,\varrho\Big]\Big\},
\qquad A=1,2.
\eey
Always when the modified propagator $\st{\trido}{\varrho}$ appears, original quantum
mechanics (characterized by $\ro\equiv 0$) of decomposed systems shows "thermodynamical
properties", here a
"heat exchange" and an "entropy exchange", a non-vanishing "entropy time rate" and not expected
an "entropy production" \R{L16b} below. These facts are sufficient to consider the modified von
Neumann equations \R{+45b}.

For the partition \#(12) between the sub-systems follows analogously to \R{AL3a}$_2$
\bey\nonumber
\st{\td}{Q}{^{12}} &:=& \mbox{Tr}({\cal H}{^{12}}\st{\td}{\varrho})\ =\
\mbox{Tr}\Big\{{\cal H}{^{12}}
\Big(\ro-\frac{i}{\hbar} \Big[{\cal H},\varrho\Big]\Big)\Big\} =\
\\ \label{aL3a}
&=&
\mbox{Tr}\Big\{{\cal H}{^{12}}
\Big(\ro-\frac{i}{\hbar} \Big[({\cal H}^1+{\cal H}^2),\varrho\Big]\Big)\Big\}.
\eey
Inserting \R{+49}, \R{AL3a} and \R{aL3a} results in external and internal heat exchanges and
in the influences of the partition between the sub-systems
\byy{MM1}
\st{\td}{Q}{^A_{ex}} &:=& \mbox{Tr}({\cal H}\!{^A\ro_{ex}})\ =\
\mbox{Tr}^A({\cal H}\!{^A\ro{_{ex}^A}}),
\\ \nonumber
\st{\td}{Q}{^A_{int}} &:=& \mbox{Tr}({\cal H}\!{^A\ro_{iso}})
+\frac{i}{\hbar}\mbox{Tr}\Big\{{\cal H}^{12}\Big[{\cal H}^A,\varrho\Big]\Big\}\ =\ 
\\ \label{MM2} &=&
\mbox{Tr}\Big\{{\cal H}\!{^A\Big(\ro_{iso}}
-\frac{i}{\hbar}\Big[{\cal H}^{12},\varrho\Big]\Big)\Big\},\quad A=1,2,
\\ \label{MM3}
\st{\td}{Q}{^{12}_{ex}} &:=& \mbox{Tr}({\cal H}{^{12}}\ro_{ex}),
\\ \label{MM4}
\st{\td}{Q}{^{12}_{int}} &:=& \mbox{Tr}\Big\{{\cal H}{^{12}}\Big(\ro_{iso}
-\frac{i}{\hbar}\Big[({\cal H}^1+{\cal H}^2),\varrho\Big]\Big)\Big\}.
\eey
From \R{AL3a}$_2$, \R{aL3a}$_1$, \R{+11}$_2$ and \R{+12}$_1$ follows the total heat
exchange, external and internal, of the system
\bee{BL3a}
\st{\td}{Q}{^{1}}+\st{\td}{Q}{^{2}}+\st{\td}{Q}{^{12}}\ =\ \st{\td}{Q}\ =\
\mbox{Tr}({\cal H}\ro).
\ee
According to \R{§1}$_2$, the heat exchanges of the decomposed described system
($\st{\td}{Q}\!{^1}, \st{\td}{Q}\!{^2}, \st{\td}{Q}\!{^{12}}$) are additive and equal to the
heat exchange \R{+12}$_1$ $\st{\td}{Q}$ of the undecomposed described system.

Splitting the propagator $\ro$ of the undecomposed system according to \R{+49} into its
exchange and irreversibility parts, the partial heat exchanges \R{AL3a} and \R{aL3a}
decompose into an external and an internal part
\byy{dL3a}
\st{\td}{Q}{^A}\ =\ \st{\td}{Q}{^A_{ex}} + \st{\td}{Q}{^A_{int}},&\quad&
\st{\td}{Q}{^{12}}\ =\ \st{\td}{Q}{^{12}_{ex}} + \st{\td}{Q}{^{12}_{int}}.
\\ \label{1dL3a}
\st{\td}{Q}{^{1}_{ex}}+\st{\td}{Q}{^{2}_{ex}}+\st{\td}{Q}{^{12}_{ex}} &=:& \st{\td}{Q}_{ex}\ =\ \mbox{Tr}({\cal H}\ro_{ex}),
\\ \label{2dL3a}
\st{\td}{Q}{^{1}_{int}}+\st{\td}{Q}{^{2}_{int}}+\st{\td}{Q}{^{12}_{int}} &=:& \st{\td}{Q}_{int}\ =\ \mbox{Tr}({\cal H}\ro_{iso}).
\eey

If the sum of the internal heat exchanges is set to zero and taking into account that the
partition has no contact to the system's environment, from \R{2dL3a}$_2$ and \R{MM3} follows
the thermodynamical setting\footnote{\R{aL3a3}$_1$ already known from \R{20a}}
\bee{aL3a3}\blacksquare\hspace{.8cm}
\st{\td}{Q}{_{int}}\ =\
\mbox{Tr}({\cal H}\ro_{iso})\ \equiv\ 0,\quad
\st{\td}{Q}{^{12}_{ex}}\ =\ \mbox{Tr}({\cal H}{^{12}}\ro_{ex})\ \equiv\ 0,
\hspace{.6cm}\blacksquare
\ee
as conditions which $\ro_{iso}$ and $\ro_{ex}$ have to satisfy.
According to \R{MM1}$_2$, \R{aL3a3}, \R{1dL3a} and \R{2dL3a} result in
\byy{1dL3ax}
\st{\td}{Q}{^{1}_{ex}}+\st{\td}{Q}{^{2}_{ex}}\ =\ \st{\td}{Q}_{ex}\ =\
\mbox{Tr}^1({\cal H}^1\ro{_{ex}^1})+\mbox{Tr}^2({\cal H}^2\ro{_{ex}^2})\ =\ 
\mbox{Tr}({\cal H}\ro_{ex}),\hspace{.2cm}
\\ \label{2dL3ax}
\st{\td}{Q}{^{1}_{int}}+\st{\td}{Q}{^{2}_{int}}+\st{\td}{Q}{^{12}_{int}}\ =\ 0\ =\
\mbox{Tr}({\cal H}\ro_{iso}),\hspace{5cm}
\eey
depicting that the external heat exchanges are additive and their sum is equal to the heat exchange
of the undecomposed system. $\st{\td}{Q}{^{A}_{int}}$ is the heat exchange corresponding
to the sub-system \#$A$, and $\st{\td}{Q}{^{12}_{int}}$ is the "non-inertness" (with respect to
the heat exchange) of the partition \#(12).

From \R{2dL3ax}$_1$ follows
\bee{aL3a6}
-\st{\td}{Q}{^1_{int}}\ =\ \st{\td}{Q}{^2_{int}}+\st{\td}{Q}{^{12}_{int}},
\ee
a relation which is analogous to \R{38e1}, if the partition between the sub-systems is not inert.
If $\st{\td}{Q}{^{12}_{int}}>0$, this partition is heat absorbing, if $\st{\td}{Q}{^{12}_{int}}<0$,
the partition is heat emitting. Consequently, $\st{\td}{Q}{^{12}_{int}}$ is not a "heat exchange",
but a heating (or cooling) of the partition which prevents continuous heat exchange through it.
$\st{\td}{Q}{^{12}_{int}}$ vanishes with ${\cal H}^{12}$ according to \R{MM4}. Consequently,
the non-inertness is based on the quantum theoretical interaction.

Taking \R{aL3a3}$_1$ into account, the non-inertness \R{MM4}
becomes
\byy{aL3a7}
\st{\td}{Q}{^{12}_{int}} &:=& -\mbox{Tr}\Big\{({\cal H}^1+{\cal H}^2)\Big(\ro_{iso}
-\frac{i}{\hbar}\Big[{\cal H}{^{12}},\varrho\Big]\Big)\Big\}\ =
\\ \label{aL3a7a}
&=&-\mbox{Tr}^1({\cal H}^1\!\ro{^1_{iso}})-\mbox{Tr}^2({\cal H}^2\!\ro{^2_{iso}})+
\mbox{Tr}\Big\{({\cal H}^1+{\cal H}^2)\frac{i}{\hbar}\Big[{\cal H}{^{12}},\varrho\Big]\Big\}.
\hspace{.5cm}
\eey
A comparison with \R{AL3a}$_4$ shows that \R{2dL3ax}$_1$ is satisfied.

In general,
the internal heat exchanges are not ccontinuous at the partition between the sub-systems \#1
and \#2, if the non-inertness $\st{\td}{Q}\!{^{12}_{int}}$ does not vanish according to \R{aL3a6}.
Consequently, an inert partition is according to \R{MM4} characterized by
\bee{MMe}
\mbox{inert:}\qquad
\st{\td}{Q}{^{12}_{int}}\ \equiv\ 0\ =\ \mbox{Tr}\Big\{{\cal H}{^{12}}\Big(\ro_{iso}
-\frac{i}{\hbar}\Big[{\cal H},\varrho\Big]\Big)\Big\},
\ee
and taking \R{aL3a3}$_2$ and \R{+6}$_1$ into account,
\bee{MMf}
\mbox{inert:}\qquad
0\ =\ \mbox{Tr}\Big\{{\cal H}{^{12}}\Big(\ro
-\frac{i}{\hbar}\Big[{\cal H},\varrho\Big]\Big)\Big\}\ =\
\mbox{Tr}\Big\{{\cal H}{^{12}}\st{\td}{\varrho}\Big\}
\ee
is a constraint for the time derivative of the density operator of processes in bipartite systems equipped with an inert partition.

Concerning the heat exchanges, the following facts are valid:
\begin{itemize}
\item According to the decomposition of the Hamiltonian of the bipartite system, there are two
heat exchanges, $\st{\td}{Q}\!{^A}$, A=1,2, and the non-inertness,
$\st{\td}{Q^{12}}$, \R{AL3a} and \R{aL3a}.
\item Inserting the decomposed propagator \R{+49}, external and internal heat exchanges are
generated, \R{MM1} to \R{MM4}.
\item As a thermodynamical input is chosen: the sum of all internal heat exchanges vanishes
\R{aL3a3}$_1$, and the partition has no external contact and therefore the corresponding external
heat exchange vanishes \R{aL3a3}$_2$.
\item The heat exchange of the undecomposed system is the sum of the external heat 
exchanges \R{1dL3ax}$_1$.
\item The internal heat exchange is not continuous at the partition, if ${\st{\td}{Q}{^{12}_{int}}}$,
the non-inertness, does not vanish. A dynamical constraint is found for processes in bipartite
systems equipped with an inert partition \R{MMf}.
\end{itemize}

In original quantum mechanics, $\ro\equiv 0$, the external heat
exchanges vanish according to \R{MM1} and \R{MM3}, that means, the considered bipartite
system is adiabatically isolated from its environment, but power exchanges may exist according to
\R{38d4}$_1$ so long as the time derivatives of the work variables $\st{\td}{a}\!{^A}$ do not
vanish. The internal heat exchanges, \R{MM2} and \R{MM4}, vanish with ${\cal H}^{12}$ in
original quantum mehanics.

\subsubsection{External heat exchanges and contact temperature}

The contact temperature $\Theta$ is defined by an inequality such as 
\R{+21} for {\bf undecomposed systems} in an equilibrium environment of the thermostatic
temperature $T^\Box$ \footnote{The temperature of the environment is written in brackets
$\st{\td}{Q}\!{_{ex}}(T^\Box)$.}
\bee{DU3a}
\Theta\ \mbox{fix,}\ T^\Box\ \mbox{variable}:\qquad
\Big(\frac{1}{\Theta}-\frac{1}{T^\Box}\Big)\st{\td}{Q}(T^\Box)\ \geq\ 0,\quad
\st{\td}{Q}\ \equiv\ \st{\td}{Q}\!{_{ex}}.
\ee
According to the definition of the contact temperature, the heat exchange vanishes with change of
sign at $T^\Box=\Theta$
\bee{DU3b}
\st{\td}{Q}\!{_{ex}}(\Theta)\ =\ 0.
\ee
The external heat exchange $\st{\td}{Q}\!_{ex}$ in \R{DU3a}
depends on the temperature difference located in the brackets left of it, described by a
constitutive equation of the shape
\bee{L27a} 
\st{\td}{Q}{_{ex}(T^\Box)} =\ \mbox{Tr}\Big({\cal H}\ro (\Theta,T^\Box)\Big)\ =\ 
\Omega^{ex}\Big(\frac{1}{\Theta}-\frac{1}{T^\Box}\Big)\ \equiv\ \Omega^{ex}(x)
\ee
according to \R{+12}$_1$.

$\blacksquare$
Presupposed that small temperature differences $x$ generate
small heat exchanges\footnote{this is a fact of experiments},
$\Omega^{ex}(x)$ is a continuous function of $x$. Also a fact of experience is that the heat exchange
is as greater as the temperature difference $x$ is
\bee{L27a1y}
x\ >\ y\ \longleftrightarrow\ \Omega^{ex}(x)\ >\ \Omega^{ex}(y),\qquad
x\ =\ y\ \longleftrightarrow\ \Omega^{ex}(x)\ =\ \Omega^{ex}(y),
\ee
that means, $\Omega^{ex}(x)$ is a strictly monotonous function.\hfill$\blacksquare$

Because of continuity and monotony of $\Omega^{ex}(x)$, the external heat exchange
has a single zero which is the environment equilibrium temperature
\byy{L27by}
\Omega^{ex}(0)\ =\ 0,\qquad\Omega^{ex}(x)\ =\ 0\ \longrightarrow\ x\ =\ 0,
\\ \label{L29x}
\st{\td}{Q}\!{_{ex}}(T^\Box)\ =\ 0\ \longleftrightarrow\ T^\Box = \Theta\hspace{1.3cm}
\eey
according to \R{L27a}, resulting according to \R{L27a1y}$_1$ in
\bee{L29x1}
\Omega^{ex}(x)\ \gtreqless\ 0\ \longleftrightarrow\ x \gtreqless\ 0.
\ee

For {\bf decomposed systems} \R{DU3a} writes
\bee{DU3c}
\Theta^A\ \mbox{fix,}\ T^\Box\ \mbox{variable}:\quad
\Big(\frac{1}{\Theta^A}-\frac{1}{T^\Box}\Big)\st{\td}{Q}\!{^A_{ex}}(T^\Box)\ \geq\ 0,\quad A=1,2,
\ee
with the thermostatic temperature $T^\Box$ of the equilibrium environment. Analo\-gous to \R{DU3b}
\bee{DU3c1}
T^\Box\ \doteq\ \Theta^A\ \longrightarrow\ \st{\td}{Q}\!{^A_{ex}}(\Theta^A)\ =\ 0
\ee
is valid, also with a change of sign. The constitutive equation of the heat exchanges are according to
\R{L27a}
\bee{DU3c2} 
\st{\td}{Q}{^A_{ex}(T^\Box)} =\ 
\Omega{^{ex}_A}\Big(\frac{1}{\Theta^A}-\frac{1}{T^\Box}\Big).
\ee
A comparison with \R{MM1}$_2$ shows that the propagators depend on
temperature\footnote{See more details in sect.\ref{PTD}}, resulting in
\byy{QU1}
\st{\td}{Q}{^A_{ex}(T^\Box)}\ =\ 
\mbox{Tr}^A\Big\{{\cal H}^A \ro{^A_{ex}}(\Theta^A;T^\Box)\Big\}
&=& \Omega{^{ex}_A}\Big(\frac{1}{\Theta^A}-\frac{1}{T^\Box}\Big),
\\ \label{QU1a}
\mbox{Tr}^A\Big\{{\cal H}^A \ro{^A_{ex}}(\Theta^A;\Theta^A)\Big\}&=&0.
\eey

Addition of the inequalities \R{DU3c} and use of \R{1dL3ax}$_1$ results in
\bee{DU4a}
\frac{\st{\td}{Q}\!{^1_{ex}}(T^\Box)}{\Theta^1}
+\frac{\st{\td}{Q}\!{^2_{ex}}(T^\Box)}{\Theta^2}
-\frac{\st{\td}{Q}\!{_{ex}}(T^\Box)}{T^\Box}\ \geq\ 0.
\ee
Two special cases of \R{DU4a} ($T^\Box\doteq\Theta^1$ and
$T^\Box\doteq\Theta^2$) are by taking \R{DU3c1} and \R{1dL3ax}$_1$ into account
\bee{DU7}
\Big(\frac{1}{\Theta^2}-\frac{1}{\Theta^1}\Big)\st{\td}{Q}\!{_{ex}}(\Theta^1)\ \geq\ 0,
\qquad
\Big(\frac{1}{\Theta^1}-\frac{1}{\Theta^2}\Big)\st{\td}{Q}\!{_{ex}}(\Theta^2)\ \geq\ 0,
\ee
from which follows
\bee{DU10}
\mbox{sign}\Big(\st{\td}{Q}\!{_{ex}}(\Theta^2)\Big)\ =\ 
-\mbox{sign}\Big(\st{\td}{Q}\!{_{ex}}(\Theta^1)\Big).
\ee
Consequently according to \R{DU3b}, the following inequalities of the external undecomposed
heat exchange are valid  
\bee{DU10a}
\st{\td}{Q}\!{_{ex}}(\Theta^2)\ \gtreqless\ (\st{\td}{Q}\!{_{ex}}(\Theta)=0)\ 
\gtreqless\ \st{\td}{Q}\!{_{ex}}(\Theta^1).
\ee
Consequently according \R{L27a1y}$_1$, the contact temperatures satisfy 
\bee{DU10a1}
\Theta^2\ \gtreqless\ \Theta\ \gtreqless\ \Theta^1.
\ee

Because there is no contact between the partition \#(12) and the system's equilibrium environment, 
$T^\Box$ cannot be used in connection with the internal heat exchanges
$\st{\td}{Q}\!\!{^A_{int}}$. That means, the defining inequalities \R{DU3c} have to be replaced
by other ones in the next section.

\subsubsection{Internal heat exchanges and contact temperature}

Because the definition of the contact temperature $\Theta^A$ of the sub-system \#A does not
depend on the special heat conduction, external or internal, the defining inequality of the contact
temperature in case of internal heat conduction is analogous to \R{DU3c}
\bee{NJ1}\blacksquare\quad
\Theta^A\ \mbox{fix,}\ T^A\ \mbox{variable}:\
\Big(\frac{1}{\Theta^A}-\frac{1}{T^A}\Big)\st{\td}{Q}\!{^A_{int}}(T^A)\ \geq\ 0,\quad A=1,2.
\hspace{.4cm}\blacksquare
\ee
Here, the thermostatic temperature $T^\Box$ of the environment is replaced by the variable $T^A$,
the replacement temperature \C{MUBE04}, so that analogously to \R{DU3c1} 
\bee{NJ2}
T^A\ \doteq\ \Theta^A\ \longrightarrow\ \st{\td}{Q}\!{^A_{int}}(\Theta^A)\ =\ 0
\ee
is valid.

$\blacksquare$\ Additionally, it is presupposed that the internal heat exchange is as the external
one \R{L27a1y} a continuous and monotonous function, a constitutive equation of the variable
$T^A$
\bee{DU10c} 
\st{\td}{Q}{_{int}^A}(T^A)\ =\
\Omega{^{int}}\Big(\frac{1}{\Theta^A}-\frac{1}{T^A}\Big),\quad A=1,2.\hspace{1.6cm}
\blacksquare\hspace{-1.6cm}
\ee
In contrast to \R{DU3c2}, here exists only one internal contact between the sub-systems \#1 and
\#2, whereas there are two external contacts between \#1 and the environment and \#2 and
the environment. In \R{DU3c}, $T^\Box$ (and $\Theta^A$) determines the heat exchange
$\st{\td}{Q}\!{^A_{ex}}(T^\Box)$, whereas in \R{NJ1}, the heat exchange
$\st{\td}{Q}{_{int}^A}(T^A)$ determines $T^A$ (besides the given $\Theta^A$).

A comparison of \R{DU10c} with \R{MM2} shows that also the propagator belonging to the
internal heat exchange depends on temperature
\bee{QU2}
\st{\td}{Q}{^A_{int}(T^A)}\ =\
\mbox{Tr}^A\Big\{{\cal H}^A \ro{^A_{iso}}(\Theta^A;T^A)\Big\}-
\frac{i}{\hbar}\mbox{Tr}\Big\{{\cal H}^{A}\Big[{\cal H}^{12},\varrho\Big]\Big\},
\ee
resulting according to \R{NJ2} in
\byy{QU2a}
\st{\td}{Q}{^A_{int}(\Theta^A)}\ =\ 
\mbox{Tr}^A\Big\{{\cal H}^A \ro{^A_{iso}}(\Theta^A;\Theta^A)\Big\}-
\frac{i}{\hbar}\mbox{Tr}\Big\{{\cal H}^A\Big[{\cal H}^{12},\varrho\Big]\Big\}\ =\ 0.
\\ \label{QU2b}
{\cal H}^{12}\doteq 0\ \longrightarrow\ 
\mbox{Tr}^A\Big\{{\cal H}^A \ro{^A_{iso}}(\Theta^A;\Theta^A)\Big\}\ =\ 0.
\eey
Inserting \R{QU2a} into \R{QU2} results in
\bee{QU2c}
\st{\td}{Q}{^A_{int}(T^A)}\ =\
\mbox{Tr}^A\Big\{{\cal H}^A\Big( \ro{^A_{iso}}(\Theta^A;T^A)
-\ro{^A_{iso}}(\Theta^A;\Theta^A)\Big)\Big\}.
\ee

From \R{aL3a6} follows the non-inertness by taking \R{DU10c} into account
\byy{DU4x2a}
\st{\td}{Q}\!{^1_{int}}(T^1)+\st{\td}{Q}\!{^2_{int}}(T^2)&=&
-\st{\td}{Q}\!{^{12}_{int}}(\Theta^1,\Theta^2;T^1,T^2),
\\ \label{aDU10c}
\Omega{^{int}}\Big(\frac{1}{\Theta^1}-\frac{1}{T^1}\Big)+
\Omega{^{int}}\Big(\frac{1}{\Theta^2}-\frac{1}{T^2}\Big) &=&
-\st{\td}{Q}{_{int}^{12}}(\Theta^1,\Theta^2;T^1,T^2).
\eey
By taking \R{DU10c} and \R{QU2c} into account, \R{aDU10c} results in
\bey\nonumber
\st{\td}{Q}{_{int}^{12}}(\Theta^1,\Theta^2;T^1,T^2) &=&
-\mbox{Tr}^1\Big\{{\cal H}^1\Big( \ro{^1_{iso}}(\Theta^1;T^1)
-\ro{^1_{iso}}(\Theta^1;\Theta^1)\Big)\Big\}-\hspace{.5cm}
\\ \label{QU4}
&&-\mbox{Tr}^2\Big\{{\cal H}^2\Big( \ro{^2_{iso}}(\Theta^2;T^2)
-\ro{^2_{iso}}(\Theta^2;\Theta^2)\Big)\Big\}.
\eey

The internal heat exchange between the two sub-systems is described by two replacement
temperatures, $T^1$ and $T^2$, according to \R{NJ1}, \R{DU10c} and \R{aDU10c}
\byy{QU14y}
\Big(\frac{1}{\Theta^1}-\frac{1}{T^{1}}\Big)\st{\td}{Q}\!{^1_{int}}(T^{1})\geq\ 0,\quad
\Big(\frac{1}{\Theta^2}-\frac{1}{T^{2}}\Big)\st{\td}{Q}\!{^2_{int}}(T^{2})\ \geq\ 0,
\hspace{.5cm}
\\ \label{QU15y}
\st{\td}{Q}\!{^1_{int}}(T^{1})\ 
=\ \Omega^{int}\Big(\frac{1}{\Theta^1}-\frac{1}{T^{1}}\Big),\
\st{\td}{Q}\!{^2_{int}}(T^{2})\ 
=\ \Omega^{int}\Big(\frac{1}{\Theta^2}-\frac{1}{T^{2}}\Big).\hspace{.3cm}
\eey
Addition of \R{QU14y}$_1$ and \R{QU14y}$_2$ results in
\bee{QU15y1}
\frac{\st{\td}{Q}\!{^1_{int}}(T^{1})}{\Theta^1}
+\frac{\st{\td}{Q}\!{^2_{int}}(T^{2})}{\Theta^2}\ 
\geq\  
\frac{\st{\td}{Q}\!{^1_{int}}(T^{1})}{T^1}
+\frac{\st{\td}{Q}\!{^2_{int}}(T^{2})}{T^2}.
\ee
By taking \R{DU4x2a} into account, this inequality yields
\bey\nonumber
\frac{\st{\td}{Q}\!{^1_{int}}(T^{1})}{\Theta^1}
+\frac{\st{\td}{Q}\!{^2_{int}}(T^{2})}{\Theta^2}\ 
\geq\hspace{7cm}
\\ \nonumber
\geq\ \frac{1}{T^1}\Big(-\st{\td}{Q}\!{^2_{int}}(T^{2})
-\st{\td}{Q}\!{^{12}_{int}}(\Theta^1,\Theta^2;T^{1},T^{2})\Big)+\hspace{2.3cm}
\\ \nonumber
+\frac{1}{T^2}\Big(-\st{\td}{Q}\!{^1_{int}}(T^{1})
-\st{\td}{Q}\!{^{12}_{int}}(\Theta^1,\Theta^2;T^{1},T^{2})\Big)\ =\hspace{.3cm}
\\ \label{aQU15y1}
=\ -\frac{\st{\td}{Q}\!{^1_{int}}(T^{1})}{T^2}-\frac{\st{\td}{Q}\!{^2_{int}}(T^{2})}{T^1}
-\Big(\frac{1}{T^1}+\frac{1}{T^2}\Big)
\st{\td}{Q}\!{^{12}_{int}}(\Theta^1,\Theta^2;T^{1},T^{2}).\hspace{.3cm}
\eey
Summing up \R{QU15y1} and \R{aQU15y1} results in
\bey\nonumber
2\Big(\frac{\st{\td}{Q}\!{^1_{int}}(T^{1})}{\Theta^1}
+\frac{\st{\td}{Q}\!{^2_{int}}(T^{2})}{\Theta^2}\Big)\ \geq\ 
\Big(\frac{1}{T^1}-\frac{1}{T^2}\Big)\Big(\st{\td}{Q}\!{^1_{int}}(T^{1})
-\st{\td}{Q}\!{^2_{int}}(T^{2})\Big)-
\\ \label{abQU15y1}
-\Big(\frac{1}{T^1}+\frac{1}{T^2}\Big)
\st{\td}{Q}\!{^{12}_{int}}(\Theta^1,\Theta^2;T^{1},T^{2}).\hspace{.4cm}
\eey
The quantum-thermal version of this inequality results from inserting of \R{QU2c} and \R{QU4} into
\R{abQU15y1}\footnote{a complex expression which is suppressed here}. The heat exchanges,
the non-inertness and the expressions in the sequel look like classical relations, but according to
\R{QU2c} and \R{QU4}, these are of quantum-thermal origin.

According to \R{abQU15y1}, four cases are possible by combination: (inert or non-inert) 
combined with ($T^1\neq T^2$ or $T^1 = T^2 =: T^{12}$). For the present, {\bf non-inert
partitions}
\begin{itemize}
\item
\bee{QU8}\fbox{$
\st{\td}{Q}\!{^{12}_{int}}(\Theta^1,\Theta^2;T^1,T^2)\ \equiv\hspace{-.4cm}/
\hspace{.4cm}0$}
\ee
are considered starting with
\begin{itemize}\item[$\Box$]
\bee{MZ1y}
\fbox{$ T^1\ \doteq\ T^2\ =:\ T^{12} $}.
\ee
By taking \R{DU4x2a} into account, \R{abQU15y1} results in
\bee{QU15y3}
\frac{\st{\td}{Q}\!{^1_{int}}(T^{12})}{\Theta^1}
+\frac{\st{\td}{Q}\!{^2_{int}}(T^{12})}{\Theta^2}\ 
\geq\ -\frac{1}{T^{12}}\st{\td}{Q}\!{^{12}_{int}}(\Theta^1,\Theta^2;T^{12},T^{12}).
\ee
The interpretation of this inequality is as follows: the non-inert partition \#(12) between the
sub-systems \#1 and \#2 has its own temperature $T^{12}$ different from the contact 
temperatures of the sub-systems, $\Theta^1$ and $\Theta^2$.
Consequently, the internal heat exchange takes place through
a third non-inert sub-system \#(12), e.g. a heat reservoir.
 
A comparison of \R{QU15y3} with \R{DU4a} shows the similarity between external and internal
heat exchanges: the temperature of the environment $T^\Box$ corresponds to the temperature
$T^{12}$ of the non-inert partition, and the non-inertness corresponds to the external heat
exchange of the undecomposed system.

\item[$\Box$]
If the contact temperatures of the sub-systems are equal, 
\bee{QU18y}
\fbox{$\Theta^1\ \doteq\ \Theta^2\ =:\ \Theta^{12}$},
\ee
\R{abQU15y1} yields by use of \R{DU4x2a}
\bey\nonumber 
0\ \geq
\Big(\frac{1}{T^1}-\frac{1}{T^2}\Big)\Big(\st{\td}{Q}\!{^1_{int}}(T^{1})
-\st{\td}{Q}\!{^2_{int}}(T^{2})\Big)+\hspace{2.8cm}
\\ \label{abQU18}
+\Big(\frac{2}{\Theta^{12}}-\frac{1}{T^1}-\frac{1}{T^2}\Big)
\st{\td}{Q}\!{^{12}_{int}}(\Theta^{12},\Theta^{12};T^{1},T^{2}).
\eey
The interpretation of this inequality is as follows: Although the sub-systems \#1 and \#2 have the
same contact temperature $\Theta^{12}$, the heat exchanges through the non-inert partition
\#(12) do not vanish according to \R{QU15y}, if $T^1\neq T^2$
\byy{aQU18y1}
\st{\td}{Q}\!{^1_{int}}(T^{1})
=\Omega^{int}\Big(\frac{1}{\Theta^{12}}-\frac{1}{T^{1}}\Big)\!\neq
\st{\td}{Q}\!{^2_{int}}(T^{2})
=\Omega^{int}\Big(\frac{1}{\Theta^{12}}-\frac{1}{T^{2}}\Big).\hspace{.3cm}
\eey
Because $T^1\neq T^2$ is presupposed, a thermostatic temperature $T^{12}$ of the partition
\#(12) cannot be defined according to \R{MZ1y}. The non-inert partition presents itself as a
"thermodynamical double-sheet" which has two sides of different temperatures $T^1$ and $T^2$.
If this statement seems to be too speculative, the case $T^1=T^2$ (the "thermodynamical
mono-sheet") is discussed below. According to \R{abQU15y1}, four kinds of partitions are possible
by combination: (inert or non-inert) combined with (mono- or double-sheet).

\item[$\Box$]
If \R{MZ1y} and \R{QU18y} are jointly valid 
\bee{QU18y2}
\fbox{$ T^1\ \doteq\ T^2\ =:\ T^{12} $}\ \wedge\ 
\fbox{$\Theta^1\ \doteq\ \Theta^2\ =:\ \Theta^{12}$},
\ee
\R{QU15y3} results in
\bee{QU18y2a}
\Big(\frac{1}{T^{12}}-\frac{1}{\Theta^{12}}\Big)
\st{\td}{Q}\!{^{12}_{int}}(\Theta^{12},\Theta^{12};T^{12},T^{12})\geq\ 0,
\ee
and according to \R{QU15y}, the internal heat exchanges are equal
\bee{aQU18y2a}
\st{\td}{Q}\!{^1_{int}}(T^{12})\ 
=\ \Omega^{int}\Big(\frac{1}{\Theta^{12}}-\frac{1}{T^{12}}\Big)\ =\ 
\st{\td}{Q}\!{^2_{int}}(T^{12}),
\ee
a fact which has the following
interpretation: the partition \#(12) is a heat reservoir of the temperature $T^{12}$ surrounded by
two sub-systems of equal contact temperature $\Theta^{12}$, and the heat exchanges between the
heat reservoir and each sub-system are equal, thus cooling or heating the sub-systems with
identical heat exchanges at both sides of \#(12). According to \R{aQU18y2a} and \R{NJ2}, these
heat exchanges vanish, if $T^{12}=\Theta^{12}$. 
\end{itemize}

After having discussed the more general case of a non-inert partition, inert partitions are now
considered, e.g. diathermal interfaces\footnote{the most usual contacts}.

\item Presupposing an {\bf inert partition},
\bee{QU8a}
\fbox{$\st{\td}{Q}\!{^{12}_{int}}(\Theta^1,\Theta^2;T^1,T^2)\ \equiv\ 0$}\ \longrightarrow\ 
\st{\td}{Q}\!{^1_{int}}(T^1)\ =\ -\st{\td}{Q}\!{^2_{int}}(T^2)
\ee
follows from \R{DU4x2a}. Consequently by taking \R{NJ2} and \R{L27by} into account,
\R{QU8a}$_2$ results in
\byy{QU9a}
T^1 \doteq \Theta^1 &\longleftrightarrow&  T^2 \doteq \Theta^2,
\\ \label{QU10}
\Omega{^{int}}\Big(\frac{1}{\Theta^1}-\frac{1}{T^1}\Big) &=&
-\Omega{^{int}}\Big(\frac{1}{\Theta^2}-\frac{1}{T^2}\Big),
\\ \label{QU10a}
\longrightarrow\ 
\frac{1}{\Theta^1}-\frac{1}{T^1}\ \gtreqless &0& \gtreqless\ 
\frac{1}{\Theta^2}-\frac{1}{T^2}
\eey
according to \R{L29x1}.

The necessity of using two variables, $T^1$ and $T^2$, can be seen from
\R{QU9a}: a single variable $T^0:= T^1\equiv T^2$ satisfies \R{QU9a} only in the
case $\Theta^1=\Theta^2$ which is not valid in general.
Dependent on the contact temperatures of the sub-systems,
$\Theta^1$ and $\Theta^2$, the variables $T^1$ and $T^2$ determining the internal heat
exchanges are connected to each other for inert partitions according to \R{QU10a}
\bee{QU12} 
T^1\ \gtreqless\ \Theta^1\ \longleftrightarrow\ T^2\ \lesseqgtr\ \Theta^2
\ee
as a necessary condition of an inert partition.

A sufficient condition for vanishing non-inertness follows from the defining inequalities \R{QU14y}
which yield by taking \R{QU8a}$_2$ into account
\bee{aQU12}
0\ \leq\ \Big(-\frac{1}{\Theta^2}+\frac{1}{T^{2}}\Big)
\Big(-\st{\td}{Q}\!{^2_{int}}(T^{2})\Big)\ =\ 
\Big(-\frac{1}{\Theta^2}+\frac{1}{T^{2}}\Big)\st{\td}{Q}\!{^1_{int}}(T^{1}),
\ee
an inequality which is set identically to \R{QU15y}$_1$, resulting in
\byy{bQU12-}
\blacksquare\hspace{1cm}
0\leq
\Big(-\frac{1}{\Theta^2}+\frac{1}{T^{2}}\Big)\st{\td}{Q}\!{^1_{int}}(T^{1})&\st{\td}{\equiv}&
\Big(\frac{1}{\Theta^1}-\frac{1}{T^{1}}\Big)\st{\td}{Q}\!{^1_{int}}(T^{1})\hspace{.8cm}
\\ \label{bQU12}
\longrightarrow\ -\frac{1}{\Theta^2}+\frac{1}{T^{2}} &=&
\frac{1}{\Theta^1}-\frac{1}{T^{1}},\hspace{1.8cm}\blacksquare
\eey
in accordance with \R{QU10a} and \R{QU12}.

Equation \R{bQU12} shows, how the variables $T^1$ and $T^2$ depend on each other for inert
partitions. Inserting \R{bQU12} into \R{QU10} results in
\byy{cQU12}
\Omega{^{int}}\Big(\frac{1}{\Theta^1}-\frac{1}{T^1}\Big) &=&
-\Omega{^{int}}\Big(\frac{1}{\Theta^2}-\frac{1}{T^2}\Big) = 
-\Omega{^{int}}\Big(-\frac{1}{\Theta^1}+\frac{1}{T^1}\Big),\hspace{.7cm}
\\ \label{dQU12}
&&\longrightarrow\ \Omega{^{int}}(x)\ =\ -\Omega{^{int}}(-x).
\eey
This odd-symmetry, derived for inert partitions, supplements the properties \R{L27a1y} to \R{L29x1}
of the constitutive equation of the internal heat exchanges in general.
\begin{itemize}\item[$\Box$]
If the contact temperatures of the sub-systems separated by an inert double-sheet partition are equal
\bee{QU18}
\fbox{$\mbox{inert:}$}\hspace{1cm}\fbox{$\Theta^1\ \doteq\ \Theta^2\ =:\ \Theta^{12}$},
\ee
\R{QU12} yields
\bee{QU12a}
T^1\ \gtreqless\ \Theta^{12}\ \gtreqless\ T^2,
\ee
and the heat exchanges are according to \R{QU10} and \R{bQU12} 
\bee{QU12b}
\Omega{^{int}}\Big(\frac{1}{\Theta^{12}}-\frac{1}{T^1}\Big)\ =\
-\Omega{^{int}}\Big(\frac{1}{\Theta^{12}}-\frac{1}{T^2}\Big)\ \neq\ 0
\ee
in contrast to \R{aQU18y1} at a non-inert partition, and \R{abQU18} results by use of \R{QU8a}$_2$
and \R{QU18} at a double-sheet inert partition  in
\bee{aQU12b}
\Big(\frac{1}{T^2}-\frac{1}{T^1}\Big)\st{\td}{Q}\!{^1_{int}}(T^1)\ \geq\ 0,\qquad
\Big(\frac{1}{T^1}-\frac{1}{T^2}\Big)\st{\td}{Q}\!{^2_{int}}(T^1)\ \geq\ 0,
\ee
in accordance with \R{abQU15y1}.
Consequently, the heat exchange between two sub-systems of equal contact temperature does not
vanish in general, even if the partition between them is inert, a fact which is different from
thermostatics. The reason for the non-vanishing of the internal heat exchange between sub-systems
of equal contact temperature is $T^1\neq T^2$, the double-sheet partition.

The inequality \R{aQU12b} has the shape of a defining inequality \R{QU14y}$_1$ in which
the contact temperature $\Theta^1$ is replaced by the heat exchange variable
$T^2$.\footnote{therefore the name "replacement temperature" for $T^A$}

\item[$\Box$]
If the partition is inert and has its own temperature $T^{12}$
\bee{QU12c}
\fbox{$\mbox{inert:}$}\hspace{1cm}\fbox{$ T^1\ \doteq\ T^2\ =:\ T^{12} $},
\ee
\R{QU12} yields
\bee{QU12c1}
\Theta^1\ \gtreqless\ T^{12}\ \gtreqless\ \Theta^2,
\ee
and the heat exchanges are according to \R{QU10} and \R{bQU12}
\bee{QU12d}
\Omega{^{int}}\Big(\frac{1}{\Theta^{1}}-\frac{1}{T^{12}}\Big)\ =\
-\Omega{^{int}}\Big(\frac{1}{\Theta^{2}}-\frac{1}{T^{12}}\Big)\ \neq\ 0.
\ee
The inequality \R{QU15y3} results for an inert mono-sheet partition by use of \R{QU8a}$_2$ in
\byy{aQU12d}
\Big(\frac{1}{\Theta^1}-\frac{1}{\Theta^2}\Big)\st{\td}{Q}\!{^1_{int}}(T^{12})\ \geq\ 0,\ 
\Big(\frac{1}{\Theta^2}-\frac{1}{\Theta^1}\Big)\st{\td}{Q}\!{^2_{int}}(T^{12})\ \geq\ 0,
\\ \label{bQU12d}
\longrightarrow
\Big(\frac{1}{\Theta^1}-\frac{1}{T^{12}}\Big)\st{\td}{Q}\!{^1_{int}}(T^{12})+
\Big(\frac{1}{\Theta^2}-\frac{1}{T^{12}}\Big)\st{\td}{Q}\!{^2_{int}}(T^{12}) \geq 0,
\eey
a sum of two defining inequalities belonging to the sub-systems \#1 and \#2 with the common heat
exchange variable $T^{12}$ of the partition.

\item[$\Box$]
If \R{QU18} and \R{QU12c} are jointly valid 
\bee{QU12d1}
\fbox{$\mbox{inert:}$}\hspace{1cm}\fbox{$ T^1\ \doteq\ T^2\ =:\ T^{12} $}\ \wedge\ 
\fbox{$\Theta^1\ \doteq\ \Theta^2\ =:\ \Theta^{12}$},
\ee
\R{QU12a} and \R{QU12c1} result in
\bee{QU12d2}
\Theta^{12}\ =\ T^{12},
\ee
and \R{QU12b} and \R{QU12d} yield
\bee{QU12d3}
\Omega{^{int}}\Big(\frac{1}{\Theta^{12}}-\frac{1}{T^{12}}\Big)\ =\
-\Omega{^{int}}\Big(\frac{1}{\Theta^{12}}-\frac{1}{T^{12}}\Big)\ =\ 0.
\ee

Consequently, the following\footnote{almost evident} proposition is valid
\begin{center}
\parbox{9cm}{{\bf Proposition:} The internal heat exchange through an inert mono-sheet partition
of temperature $T^{12}$ which separates two non-equilibrium sub-systems of equal contact temperature $\Theta^{12}$ vanishes with $\Theta^{12}=T^{12}$.}
\end{center}

The contact temperature effects with respect to the heat exchange are different
from those of the thermostatic temperature: the heat exchange between two non-equilibrium
sub-systems of equal contact temperature does not vanish in general according to \R{QU12b},
even if the partition between
the sub-systems is inert\footnote{If the partition is not inert, the non-inertness prevents from
vanishing of the heat exchanges according to \R{QU18y2a} and \R{abQU18}.}, except the
mono-sheet partition's temperature is equal to the joint contact temperature of the sub-systems
according to \R{QU12d3}.

\end{itemize}

\item
In {\bf equilibrium} follows for arbitrary partitions according to \R{NJ2} and the fact that the contact
temperature changes into the thermostatic one which is equal in the two sub-systems according to
equilibrium
\bee{QU12e}
\st{\td}{Q}{^\alpha_{eq}}(T^\alpha)\ =\ 0\ \longrightarrow\ T^\alpha =
\Theta{^\alpha_{eq}} \equiv T{^\alpha_{eq}} = T{_{eq}},\qquad\alpha=1,2.
\ee

\item Some {\bf general remarks}

If the von Neumann equation is modified by the propagator $\ro$, power and heat exchanges,
\R{+11}$_1$ and \R{+12}, are introduced. Because the propagator decomposes into an exchange
part
and an irreversible part, $\ro _{ex}$ and $\ro_{iso}$ according to \R{+49}, also the power and the
heat exchanges decompose into ex- and int-parts. The decomposition of the power exchanges is
achieved by the work variables, \R{38d4} and \R{38d5}, whereas that of the heat exchanges 
is done by the parts of the
decomposed propagator, \R{MM1} to \R{MM4}. For undecomposed and bipartite
systems, non-equilibrium contact temperatures are introduced,  \R{DU3a}, \R{QU1} and \R{NJ1},
which appear in the quantum-thermal representation of the heat exchanges, \R{QU1} and \R{QU2}.
The heat exchanges are restricted by strictly monotonous constitutive equations, \R{L27a}, \R{QU1}
and \R{DU10c}. Thus, the propagators in the quantum-thermal representation of the heat exchanges
are connected with their corresponding classical constitutive equations. This connection has to
be taken into account by choosing the propagator of the modified von Neumann equation
\R{+6}$_1$.

\end{itemize}

After the discussion of the heat exchanges, that of the entropy exchanges follows in the next section.

\subsubsection{Entropy exchanges\label{EEX}}

For {\bf undecomposed systems}, the entropy exchange is defined by heat exchange over contact
temperature of the corresponding non-equilibrium sub-system according to \R{+18}$_1$.
Presupposing that the heat exchange between system and environment takes place through an
inert partition resulting in $\st{\td}{Q}=-\st{\td}{Q}\!{^\Box}$,
the defining inequality \R{+21} becomes
\bee{EX1}
\Xi\ =\ \frac{\st{\td}{Q}}{\Theta}
\geq\ \frac{\st{\td}{Q}}{T^\Box}\ =\ -\frac{\st{\td}{Q}\!{^\Box}}{T^\Box}\ =:\ -\Xi^\Box.
\ee
The entropy exchange $\Xi^\Box$ is bound up with the equilibrium environment of temperature
$T^\Box$, whereas $\Xi$ belongs to the undecomposed described non-equilibrium system of
contact temperature $\Theta$. Consequently by introducing the non-equilibrium
contact temperature, two different definitions of entropy exchange appear: the original one
$\Xi^\Box$ bound up with the equilibrium environment, and $\Xi$ bound up with the
non-equilibrium system. The inequality \R{EX1} shows that the amounts of these entropy
exchanges are different. The quantum-thermal version of \R{EX1} is generated by inserting
\R{+12}$_1$.

For {\bf bipartite systems}, the entropy exchange is as the heat exchange decomposed into an
external and an internal part according to \R{QU1}$_1$ and \R{QU2c}. The external part is
\bee{IM8}
\Xi{^A_{ex}}(T^\Box)\ :=\ \frac{\st{\td}{Q}{^A_{ex}}(T^\Box)}{\Theta^A}\ =\
\mbox{Tr}^A\Big(\frac{{\cal H}^A}{\Theta^A} \ro{^A_{ex}}\!(\Theta ^A;T^\Box)\Big),
\quad A=1,2.
\ee
and the inequalites \R{DU3a} and \R{DU4a} transform into
\byy{EX2b}
\Xi_{ex}(T^\Box)\ :=\ \frac{\st{\td}{Q}_{ex}\!(T^\Box)}{\Theta}
&\geq& \frac{\st{\td}{Q}_{ex}\!(T^\Box)}{T^\Box}\ =\ 
-\frac{\st{\td}{Q}\!{^\Box_{ex}}(T^\Box)}{T^\Box}\ =:\ -\Xi{^\Box_{ex}}(T^\Box),
\hspace{.7cm}
\\ \label{EX2}
\Xi{^1_{ex}}(T^\Box)+\Xi{^2_{ex}}(T^\Box) &\geq& -\Xi{_{ex}^\Box}(T^\Box).
\eey
The sum of the entropy exchanges of the decomposed system (LHS of \R{EX2}) is greater than
the original entropy exchange bound up with the equilibrium environment (RHS of \R{EX2}).

The inequality of the internal entropy exchanges is charged with the non-inertness which remains
out of play for the external part. The inequality \R{abQU15y1} results in
\bey\nonumber
\Xi{^1_{int}}(T^1)+\Xi{^2_{int}}(T^2)\ \geq
\frac{1}{2}\Big(\frac{1}{T^1}-\frac{1}{T^2}\Big)\Big(\st{\td}{Q}\!{^1_{int}}(T^{1})
-\st{\td}{Q}\!{^2_{int}}(T^{2})\Big)-
\\ \label{EX§1}
-\frac{1}{2}\Big(\frac{1}{T^1}+\frac{1}{T^2}\Big)
\st{\td}{Q}\!{^{12}_{int}}(\Theta^1,\Theta^2;T^{1},T^{2}),\hspace{.8cm}
\eey
and the internal entropy exchanges are according to \R{QU2c} and \R{QU2} defined by
\bey\nonumber
\Xi{^A_{int}}(T^A) &:=&
\mbox{Tr}^A\Big\{\frac{{\cal H}^A}{\Theta^A}\Big( \ro{^A_{iso}}(\Theta^A;T^A)
-\ro{^A_{iso}}(\Theta^A;\Theta^A)\Big)\Big\}=
\\ \nonumber
&=&
\mbox{Tr}^A\Big\{\frac{{\cal H}^A}{\Theta^A} \ro{^A_{iso}}(\Theta^A;T^A)\Big\}-
\frac{i}{\hbar}\mbox{Tr}\Big\{\frac{{\cal H}^{A}}{\Theta^A}
\Big[{\cal H}^{12},\varrho\Big]\Big\},
\\  \label{EX§1a}
&=&
\mbox{Tr}\Big\{\frac{{\cal H}^A}{\Theta^A}\Big( \ro{_{iso}}-
\frac{i}{\hbar}\Big[{\cal H}^{12},\varrho\Big]\Big)\Big\}.
\eey

The inequality \R{EX§1} takes different shapes depending on the special partition between the
sub-systems
\#1 and \#2:
\byy{EX§2}
\mbox{non-inert, double-sheet:}\hspace{-.3cm} &&\st{\td}{Q}\!{^{12}_{int}}(\Theta^1,\Theta^2;T^{1},T^{2})
\equiv\hspace{-.4cm}/\hspace{.4cm}0,\ T^1 \neq T^2,
\\ \label{EX§3}
 \mbox{non-inert, mono-sheet:}\hspace{-.3cm} &&\st{\td}{Q}\!{^{12}_{int}}(\Theta^1,\Theta^2;T^{12},T^{12})
\equiv\hspace{-.4cm}/\hspace{.4cm}0,\ T^1 = T^2 =:T^{12},\hspace{.9cm}
\\ \label{EX§4}
\mbox{inert, double-sheet:}\hspace{-.3cm} &&\st{\td}{Q}\!{^{12}_{int}}(\Theta^1,\Theta^2;T^{1},T^{2})
\equiv 0,\ T^1 \neq T^2,
\\ \label{EX§5}
\mbox{inert, mono-sheet:}\hspace{-.3cm} &&\st{\td}{Q}\!{^{12}_{int}}(\Theta^1,\Theta^2;T^{12},T^{12})
\equiv 0,\ T^1 = T^2 =:T^{12}.
\eey

Taking \R{QU8a}$_2$ into account, \R{EX§1} becomes by use of \R{EX§4} and \R{EX§3}
\byy{EX§6}
\Xi{^1_{int}}(T^1)+\Xi{^2_{int}}(T^2) &\geq&
\Big(\frac{1}{T^1}-\frac{1}{T^2}\Big)\st{\td}{Q}\!{^1_{int}}(T^{1}),
\\ \label{EX§6a}
\Xi{^1_{int}}(T^{12})+\Xi{^2_{int}}(T^{12}) &\geq&
-\frac{1}{T^{12}}\st{\td}{Q}\!{^{12}_{int}}(\Theta^1,\Theta^2;T^{12},T^{12}),
\eey
or for the inert mono-sheet partition \R{EX§5}
\bee{EX§7}
\Xi{^1_{int}}(T^{12})+\Xi{^2_{int}}(T^{12})\ \geq\ 0.
\ee

The inequalities \R{EX2} and \R{EX§1} are quantum-thermal ones: the entropy exchanges
\R{IM8}$_1$, \R{EX2b}$_1$ and \R{EX§1a} are defined by using
$\st{\td}{Q}{^A_{ex}}(T^\Box),$ ${\st{\td}{Q}_{ex}\!(T^\Box)}$,\
${\st{\td}{Q}{^A_{int}}(T^A)}$ and 
$\st{\td}{Q}\!{^{12}_{int}}(\Theta^1\Theta^2;T^{1},T^{2})$ which on their part
are quantum-thermally defined by \R{QU1}$_1$,\R{1dL3ax}, \R{QU2c} and \R{QU4}.

In contrast to the additivity of the heat exchanges according to \R{1dL3ax}$_1$ and \R{DU4x2a},
the entropy exchanges are not additive in general according to \R{EX2} and \R{EX§1}.
Especially, the external entropy exchange of the undecomposed system is not equal
to the sum of the external entropy exchanges of the decomposed system according to \R{EX2}.
This fact is called the {\em compound deficiency} of the entropy exchanges \C{MUBE04,MUBE07}.
Compound deficiencies will be treated in more detail in sect.\ref{CD}.

A comparison of \R{38d4} and \R{38d5} with \R{MM1} and \R{MM2} depicts
that the decomposition into the external and internal parts is different for power-,
heat- and entropy-exchanges. Whereas this decomposition is achieved for the power
exchanges by the work variables,
the heat- and entropy-exchanges are decomposed by the decomposition of the propagator.
Essential is that the sum of the external entropy exchanges of the sub-systems is not equal to
the entropy exchange of the corresponding undecomposed system according to \R{EX2}.

\subsection{Propagator's temperature dependence\label{PTD}}

Because the heat exchanges \R{QU1} and \R{QU2} depend on the contact temperature of the
sub-system and on an additional variable, and because the heat exchanges are represented by a
quantum-thermal expression which contains the corresponding propagator, also this propagator
depends on the used temperatures. Now, the propagators of the different description of
the systems, undecomposed or decomposed, are connected by tracing which influences the
domain of the propagators spanned by the temperatures as demonstrated in this section:

According to \R{+21}, the heat exchange of an undecomposed system depends on the contact
temperature $\Theta$ and on the temperature of the environment $T^\Box$. Consequently
according to \R{+12}$_1$, the temperature dependence of the propagator is for an
\bee{TT1}
\mbox{undecomposed system:}\qquad\quad\ro (\Theta,T^\Box).
\ee

More specific is the situation of a
\bee{TT2}
\mbox{bipartite system:}\qquad\quad\ro (\Theta,\Theta^1,\Theta^2,T^\Box,T^1,T^2)
\ee
whose heat exchanges depend on the temperature of the environment $T^\Box$, on the contact
temperatures of the undecomposed system $\Theta$ and those of the two sub-systems $\Theta^A$ and on the heat exchange variables $T^A$ according to \R{DU3c2} and \R{DU10c}. The
decomposition of the propagator \R{+49} results in
\bey\nonumber 
\ro (\Theta,\Theta^1,\Theta^2,T^\Box,T^1,T^2)\ =\ \hspace{5cm}
\\ \label{TT3} 
=\ \ro_{ex}(\Theta,\Theta^1,\Theta^2,T^\Box)+\ro_{iso}(\Theta^1,\Theta^2,T^1,T^2),
\hspace{-.8cm}
\eey
whereas the tracing yields
\bey\nonumber 
\ro (\Theta,\Theta^1,\Theta^2,T^\Box,T^1,T^2)\ &\longrightarrow&
\mbox{Tr}^2\!\ro\ =\ \ro{^1},\quad\mbox{Tr}^1\!\ro\ =\ \ro{^2},
\\ \label{TT4} 
&\longrightarrow& \ro{^1}(\Theta^1,T^\Box,T^1),\quad\ro{^2}(\Theta^2,T^\Box,T^2),
\hspace{.8cm}
\eey
demonstrating that this tracing changes also the domain of the propagators.
Thus, \R{TT3} results by tracing in
\bee{TT5}
\ro{_{ex}^1}(\Theta^1,T^\Box),\quad\ro{_{ex}^2}(\Theta^2,T^\Box),\quad
\ro{_{iso}^1}(\Theta^1,T^1),\quad\ro{_{iso}^2}(\Theta^2,T^2).
\ee
The same result appears, if the propagators of \R{TT4} are decomposed into their ex- and iso-parts
\byy{TT6}
\ro{^1}(\Theta^1,T^\Box,T^1) &=&
\ro{_{ex}^1}(\Theta^1,T^\Box)+\ro{_{iso}^1}(\Theta^1,T^1),
\\ \label{TT6a}
\ro{^2}(\Theta^2,T^\Box,T^2) &=& 
\ro{_{ex}^2}(\Theta^2,T^\Box)+\ro{_{iso}^2}(\Theta^2,T^2).
\eey
Decomposition (dec) and tracing (tra) of propagators represent a closed symmetric diagram:
\bee{TT7} 
\R{TT5}\ \st{{dec}}{\longleftarrow}\ \R{TT4}\ \st{{tra}}{\longleftarrow}\ \R{TT2}\ 
\st{{dec}}{\longrightarrow}\ \R{TT3}\ \st{{tra}}{\longrightarrow}\ \R{TT5}.
\ee

\subsection{First Law}

The time rate of the energy is determined by those of the power- and heat exchanges. According to
\R{+9}$_2$, we define the energy time rate of sub-system \#A
\bee{=1}
\st{\td}{E}\!{^A}\ :=\ \st{\td}{W}\!{^A_{ex}}+\st{\td}{W}\!{^A_{int}}+\st{\td}{Q}\!{^A_{ex}}
+\st{\td}{Q}\!{^A_{int}}.
\ee
The terms including ${\cal H}^{12}$ result in the rate of the interaction energy according to
\R{38e1} and \R{aL3a6}
\bee{=2}
\st{\td}{E}\!{^{12}}\ :=\ \st{\td}{W}{^{12}_{int}}+\st{\td}{Q}{^{12}_{int}}\ =\ 
-\st{\td}{W}\!{^1_{int}}-\st{\td}{W}\!{^2_{int}}
-\st{\td}{Q}\!{^1_{int}}-\st{\td}{Q}\!{^2_{int}},
\ee
resulting in
\bee{=3}
\st{\td}{E}\!{^{1}}+\st{\td}{E}\!{^{2}}+\st{\td}{E}\!{^{12}}\ =\  
\st{\td}{W}_{ex}+\st{\td}{Q}_{ex}\ =\ \st{\td}{E}
\ee
according to \R{38d6}$_1$ and \R{1dL3ax}$_1$. As expected, the sum of the energy changes of
the sub-sytems $(\st{\td}{E}\!{^{1}}+\st{\td}{E}\!{^{2}})$ is different from the energy change
$\st{\td}{E}$ of the undecomposed system, a fact which is called {\em compound deficiency}
(see sect.\ref{CD}) and which is caused by the quantum-theoretical interaction.

This clear formulation of the First Law indicates that the exchange quantities power- and heat
exchange are suitably defined in sects.\ref{PEX} and \ref{HEX}.

\subsection{Entropy}
\subsubsection{Partial entropies\label{PE}}

Because in decomposed systems, the density operator $\varrho$ of the undecomposed
system is replaced by those of the sub-systems of the bipartite system, \R{§2} and \R{§2a},
we are able to define partial entropies of the sub-systems.
Starting with the Shannon entropy \C{SH48,GOPE81} of the undecomposed system \R{+14}
\bee{+111}
S(\varrho)\ =\ -k_B\mbox{Tr}(\varrho\ln\varrho)
\ee
the partial entropies of the sub-systems are defined by
\bee{+112}
S_{1}(\varrho^{1})\ :=\ 
-k_B\mbox{Tr}^1(\varrho^{1}\ln\varrho^{1}),\qquad
S_{2}(\varrho^{2})\ :=\ -k_B\mbox{Tr}^2(\varrho^{2}\ln\varrho^{2})
\ee
using the partial density operators \R{§2} and \R{§2a}.

Entropy is by definition a state function.
Consequently, it contains only parameters of the system to which the entropy belongs. The
Shannon entropies \R{+111} and \R{+112} are such state functions of the undecomposed
system and of its sub-systems \#1 and \#2 \footnote{The non-inertness \R{MM4}
of the partition is not taken into the definition of entropy.}. There are a lot of different definitions
of entropy \C{RE61,TS09,STR1} which all result in quantum thermodynamics of different shape,
because the entropy production depends on the defined entropy. In contrast to the measurable
quantities, such as power, heat exchange, contact temperature, there is no measuring instrument
for entropy which comes as a definition into play generating an unequivocal thermodynamical
structure.

According to \R{3.3} in sect.\ref{TRA}, the partial entropies \R{+112} result in
\bee{+113}
S_{1}(\varrho^{1})\ =\ -k_B\mbox{Tr}(\varrho\ln\varrho^{1}),\qquad
S_{2}(\varrho^{2})\ =\ -k_B\mbox{Tr}(\varrho\ln\varrho^{2}).
\ee
A comparison of \R{+113} with \R{+111} depicts that the partial entropies are not additive with respect to the entropy of the undecomposed system
\bee{+114}
S_1(\varrho^1) + S_2(\varrho^2) - S(\varrho)\ =\ -k_B\mbox{Tr}\Big\{\varrho\Big(\ln(\varrho^{1}
\varrho^{2})-\ln\varrho\Big)\Big\},
\ee
because the density operator of the undecomposed system does not decompose in general.

Using Klein's inequality \C{KLEIN}
\bee{+115}
\mbox{Tr}(A\ln B)-\mbox{Tr}(A\ln A)\ \leq\ \mbox{Tr}B - \mbox{Tr}A, 
\ee
we obtain according to \R{+115} and \R{a61}
\byy{+116}
\mbox{Tr}\Big\{\varrho\Big(\ln(\varrho^{1}\varrho^{2})
-\ln\varrho\Big)\Big\}\ \leq\ \mbox{Tr}(\varrho^{1}\varrho^{2})
-\mbox{Tr}\varrho\ =\ 0,
\\ \label{+116a}
\mbox{Tr}(\varrho^{1}\varrho^{2})\ = \
\mbox{Tr}^1\mbox{Tr}^2(\varrho^{1}\varrho^{2})\ =\ 1,\hspace{1cm}
\eey
and \R{+114} results in
\bee{+117}
S_1(\varrho^1) + S_2(\varrho^2)\ \geq\ S(\varrho).
\ee
If the entropy $S$ of the undecomposed system does not decompose into the partial entropies of
the decomposed system, it is according to \R{+117} smaller than the sum of the partial entropies.
From \R{+114} follows
\bee{+117a}
\varrho\ \doteq\ \varrho^1\varrho^2\ \longrightarrow\
S_1(\varrho^1) + S_2(\varrho^2)\ =\ S(\varrho^1\varrho^2),
\ee
and the entropies of the sub-systems are additive in case of a decomposed statistical operator of
the undecomposed system.

\subsubsection{Entropy rates}

Starting with \R{+111} and \R{+112}
\bee{L4}
S_A\ =\ -k_B\mbox{Tr}(\varrho\ln\varrho^A)\ =\
-k_B\mbox{Tr}^A(\varrho^A\ln\varrho^A)
\ee
results in by taking \R{+44} into account
\bey\nonumber
\st{\td}{S}_A &=& -k_B\mbox{Tr}^A(\st{\td}{\varrho}{^A}\ln(Z\varrho^A))\ =\
\\ \nonumber
 &=&
-k_B\mbox{Tr}^A\Big\{\Big(-\frac{i}{\hbar}\Big[{\cal H}^A,\varrho^A\Big]
-\frac{i}{\hbar}\mbox{Tr}^B\Big[{\cal H}^{12},\varrho\Big]+\ro{^A}\Big)
\ln(Z\varrho^A)\Big\}=\hspace{1cm}
\\ \nonumber
&=&-k_B\mbox{Tr}^A\Big\{\Big(-\frac{i}{\hbar}\mbox{Tr}^B\Big[{\cal H}^{12},\varrho\Big]+\ro{^A}\Big)\ln(Z\varrho^A)\Big\}\ =\
\\ \label{L5}
&=& -k_B\mbox{Tr}\Big\{\Big(\ro-\frac{i}{\hbar}\Big[{\cal H}^{12},\varrho\Big]\Big)\ln(Z\varrho^A)\Big\},\quad A,B=1,2; A\neq B.
\eey
Summing up the entropy time rates of the sub-systems,
we obtain from \R{L5}$_4$ and \R{+15}
\bey\nonumber
&&\st{\td}{S}_1+\st{\td}{S}_2 - \st{\td}{S}\ =\
\\ \label{+131}
&&=\ -k_B\mbox{Tr}\Big\{\Big(\ro-\frac{i}{\hbar}\Big[{\cal H}^{12},\varrho\Big]\Big)\ln(Z^1\varrho^1Z^2\varrho^2)-\ro\ln(Z\varrho)\Big\},
\eey
that means, the sum of the entropy rates of the sub-sytems is different from the entropy
rate of the undecomposed system. The same non-additivity which appears for the entropy
itself according to \R{+117} is called {\em compound deficiency} (sect.\ref{CD}).

If the density operator $\varrho$ of the undecomposed system decomposes, from
\R{+114} and \R{+131} the additivity of entropy and entropy time rate is obtained
\byy{131a}
\varrho\ \doteq\ \varrho^1\varrho^2,\quad Z\ \doteq\ Z^1Z^2  
&\longrightarrow& S\ =\ S_1 + S_2,
\\ \label{a131}
\mbox{and additional}\ {\cal H}^{12}\doteq \underline{0} &\longrightarrow&
\st{\td}{S}\ =\ \st{\td}{S}_1 + \st{\td}{S}_2.
\eey

Introducing the decomposition \R{+49}, the entropy time rate \R{L5}$_4$ splits into an
external and an internal part
\byy{L5A}
\st{\td}{S}{_A^{ex}} &:=& -k_B\mbox{Tr}\Big\{\ro_{ex}\ln(Z\varrho^A)\Big\},
\\ \label{L5B}
\st{\td}{S}{_A^{int}} &:=&  -k_B\mbox{Tr}\Big\{\Big(\ro_{iso}-\frac{i}{\hbar}\Big[{\cal H}^{12},\varrho\Big]\Big)\ln(Z\varrho^A)\Big\}.
\eey

\subsubsection{Separation Axiom\label{SAX}}

According to \R{+117a} and \R{131a}, entropy and entropy time rate are additive, if the statistical
operator decomposes and if in the case of entropy time rates the interaction operator vanishes
${\cal H}{^{12}}\doteq \underline{0}$ according to \R{+131}. Then, according to \R{+45a} and
\R{+45b}, the two sub-systems are separated
from each other. Consequently, the heat exchanges $\st{\td}{Q}\!{^A}$ vanish, whereas the
entropies $S_A$ and their time rates $\st{\td}{S}_A$ are untouched by the separation of the
sub-systems. Thus, according to \R{AL3a}$_6$ and \R{L5}$_5$
\byy{aL5}\blacksquare\hspace{2cm}
\st{\td}{Q}\!{^A_{sep}}&\doteq& \mbox{Tr}^A({\cal H}^A\ro\!{^A_{sep}})\ =\ 0,
\hspace{4cm}
\\ \label{bL5}
\st{\td}{S}{_A^{sep}}&\doteq& -k_B\mbox{Tr}^A\{\ro\!{^A_{sep}}\ln(Z\varrho^A)\}\ \neq\ 0
\hspace{2cm}\blacksquare
\eey
is valid.

As in sect.\ref{SA} is demonstrated, the following propagator is sufficient for \R{aL5} and \R{bL5}
\bee{cL5}\blacksquare\hspace{3cm}
\ro\!{^A_{sep}}\ =\ \Big[{\cal H}^A,\Big[\ln(Z\varrho{^A}),{\cal H}^A\Big]\Big].
\hspace{3cm}\blacksquare
\ee
The entropy time rate results in
\bee{dL5}
\st{\td}{S}{_A^{sep}}\ =\
-k_B\mbox{Tr}^A\Big\{\Big[\ln(Z\varrho{^A}),{\cal H}^A\Big]
\Big[\ln(Z\varrho{^A}),{\cal H}^A\Big]\Big\}\ \neq\ 0.
\ee
Because of \R{aL5}$_2$, the entropy production of \#A is equal to the corresponding entropy
time rate (see next sect. \R{L5a})
\bee{eL5}
\Sigma{_A^{sep}}\ \equiv\ \st{\td}{S}{_A^{sep}}.
\ee
These internal entropy production is unknown in original quantum mechanics because of
$\ro\equiv 0$.

\subsection{Entropy productions}

According to \R{+20}$_1$, the entropy production of the sub-system $\#A$ of a bipartite system
is defined by
\bee{L5a}
\Sigma_ A\ :=\ \st{\td}{S}_A - \Xi^A\ =\ \st{\td}{S}{_A^{ex}}+ \st{\td}{S}{_A^{int}}
-\Xi^A_{ex}-\Xi^{A}_{int}.
\ee
The second equality follows from \R{+49}, the decomposition of the propagator into its exchange
and its irreversibility part.
Because in undecomposed systems the entropy production is defined
as the time rate of entropy in isolated systems \R{?1}$_1$, the same is demanded for external
isolated sub-systems, resulting according to \R{L5A} and \R{L5B}, \R{IM8} and \R{EX§1a} in
\bey\nonumber \blacksquare\hspace{12.2cm}
\\ \nonumber
\Sigma_A\ \doteq\ \st{\td}{S}{_A^{int}}-\Xi^{A}_{int} =
-k_B\mbox{Tr}\Big\{\Big(\ro_{iso}-\frac{i}{\hbar}\Big[{\cal H}^{12},\varrho\Big]\Big)\ln(Z\varrho^A)\Big\}-\hspace{1.4cm}
\\ \nonumber
-\mbox{Tr}\Big\{\frac{{\cal H}^A}{\Theta^A}\Big(\ro{_{iso}}-
\frac{i}{\hbar}\Big[{\cal H}^{12},\varrho\Big]\Big)\Big\}=\hspace{2.4cm}
\\ \label{L5c}
= -\mbox{Tr}\Big\{\Big(\frac{{\cal H}^A}{\Theta^A}+k_B\ln(Z\varrho^A)\Big) \Big(\ro_{iso}
-\frac{i}{\hbar} \Big[{\cal H}^{12},\varrho\Big]\Big)\Big\},
\\ \nonumber
0\ \doteq\ \st{\td}{S}{_A^{ex}}-\Xi^A_{ex}\ =\ -
 k_B\mbox{Tr}\Big\{\ro_{ex}\ln(Z\varrho^A)\Big\}-\mbox{Tr}\Big(\frac{{\cal H}^A}{\Theta^A} \ro{_{ex}}\Big)=\hspace{.9cm}
\\ \label{L5b}
=\ -\mbox{Tr}\Big\{\Big(\frac{{\cal H}^A}{\Theta^A}+k_B\ln(Z\varrho^A)\Big)\ro_{ex}\Big\}.
\hspace{2.2cm}\blacksquare
\eey

This expression corresponds to \R{c?1} and yields the contact temperature of sub-system \#A
in quantum-theoretical formulation
\bee{c?1ay}
\frac{1}{\Theta^A}\ =\ \frac{-\mbox{Tr}\Big\{k_B\ln(Z\varrho^A)\ro_{ex}\Big\}}
{\mbox{Tr}\{{\cal H}^A\ro_{ex}\}}.
\ee
Adding \R{L5b} and \R{L5c} results in the partial entropy production of sub-system \#A
\bee{L5c1}
\Sigma_A\ =\ 
-\mbox{Tr}\Big\{\Big(\frac{{\cal H}^A}{\Theta^A}+k_B\ln(Z\varrho^A)\Big) \Big(\ro
-\frac{i}{\hbar} \Big[{\cal H}^{12},\varrho\Big]\Big)\Big\}.
\ee

If the two sub-systems are separated from each other, ${\cal H}^{12}\doteq \underline{0},$
\bee{L5c2}
\Sigma{_A^0}\ =\ 
-\mbox{Tr}^A\Big\{\Big(\frac{{\cal H}^A}{\Theta^A}+k_B\ln(Z\varrho^A)\Big)\ro{^A}\Big\}\ 
\st{.}{\geq}\ 0,
\ee
their entropy production is not negative and \R{L5c1} results in
\bee{L5c3}
\Sigma_A\ \geq\ 
\mbox{Tr}\Big\{\Big(\frac{{\cal H}^A}{\Theta^A}+k_B\ln(Z\varrho^A)\Big)
\frac{i}{\hbar} \Big[{\cal H}^{12},\varrho\Big]\Big)\Big\}\ =:\ \Sigma{_A^{oqu}}.
\ee
Consequently, the entropy production $\Sigma_{A}$ of \#A in quantum thermodynamics is not
smaller than $\Sigma{_A^{oqu}}$ that in original quantum mechanics characterized by
$\ro\equiv0$.

According to \R{+45d}$_2$ and \R{d?1}$_2$, the partial entropy production \R{L5c1} vani\-shes
$\Sigma{_A^{eq}}=0$ in equilibrium, as well as $\Sigma^{0eq}_A$ does for the separated
sub-system \#A in \R{L5c2}. Because of this separation, the statistical operator $\varrho^A_{eq}$
is canonical according to \R{52d}
\bee{L5c4}
\Big(\frac{{\cal H}^A}{\Theta^A}+k_B\ln(Z\varrho^A)\Big)^{eq}\ =\ \underline{0}.
\ee
Consequently, both brackets of \R{L5c1} vanish in equilibrium, and \R{L5c3} becomes
\bee{L5c5}
\Sigma_A^{eq}\ =\ 0\ \geq\ 
\mbox{Tr}\Big\{\Big(\frac{{\cal H}^A}{\Theta^A}+k_B\ln(Z\varrho^A)\Big)^{eq}
\frac{i}{\hbar} \Big[{\cal H}^{12},\varrho\Big]^{eq}\Big)\Big\}\ =\ 0\ =\ \Sigma{_A^{oqu}}.
\ee
As expected, the entropy production vanishes in equilibrium in quantum thermodynamics as well
as in original quantum mechanics.

\subsection{Original quantum mechanics and bipartite systems}

Original quantum mechanics is characterized by $\ro\equiv0$. Consequently, the discrimination
into isolated and non-isolated systems disappears according to \R{+47}, $\ro_{iso}=\ro_{ex}\equiv0$ which does not influence the power exchanges (sect.\ref{PEX}).

If the system is undecomposed, the heat exchange, the entropy time rate and the entropy
production vanish in original quantum mechanics according to sect.\ref{US}. This is the reason
why original quantum mechanics is characterized as a theory for adiabatically isolated reversible
systems.

This characterization changes for bipartite systems. According to \R{MM1} to \R{MM4}, the
external heat exchanges vanish likewise as in original quantum mechanics of undecomposed
systems (adiabatical isolation), but the internal heat exchanges do not according to \R{QU2}.
Consequently, also entropy exchange and entropy production are not identical to zero according to
\R{EX§1a} and \R{L5c}.

A shortcoming is that the quantum-theoretical definition of the partial
contact temperature \R{c?1ay} cannot be given in original quantum mechanics. If the classical
definition of contact temperature through the defining inequality \R{NJ1} is not accepted, entropy
exchange and consequently entropy production cannot be defined in original quantum mechanics although heat exchanges are existing. Consequently, the introduction of the
propagator is necessary for establishing a consistent quantum thermodynamics.

In original quantum mechanics of decomposed systems, the thermodynamical quantities heat
exchange, entropy exchange, entropy time rate and entropy production vanish with the interaction
${\cal H}^{12}$ between the sub-systems \#1 and \#2. This result was expected because with
${\cal H}^{12}\doteq\underline{0}$, the bipartite system decomposes into two undecomposed
systems.

The entropy production in original quantum mechanics \R{L5c3}
vani\-shes with ${\cal H}^{12}$, that means, $\Sigma{_A^{oqu}}$ is generated by the
contact between \#1 and \#2. If the contact is cancelled, ${\cal H}^{12}\doteq \underline{0}$, both
sub-systems are reversible $\Sigma{_A^{oqu0}}=0$, that means, entropy production is generated
by interaction of reversible systems. Thus, the statement is valid: Original quantum mechanics
of bipartite closed Schottky systems is an endoreversible theory.

The corresponding statement in quantum thermodynamics is different: Presupposing 
${\cal H}^{12}\doteq \underline{0}$, \R{L5c1} results in
\bee{L5e3}
\Sigma{_A^0}\ =\ 
-\mbox{Tr}^A\Big\{\Big(\frac{{\cal H}^A}{\Theta^A}+k_B\ln(Z\varrho^A)\Big)\ro{^A}\Big\},
\ee
that means, different from original quantum mechanics, the sub-systems themselves are not
reversible, if the contact between them is cancelled. But in equilibrium, the entropy production
vanishes in quantum thermodynamics as well as in original quantum mechanics according to
\R{L5c5}.

\subsection{Compound deficiency\label{CD}}

If an undecomposed system is divided into a bipartite system, the generated partial quantities of the
sub-systems may be or may be not additive. Not additive means, that sum up quantities of the
sub-systems do not result in the corresponding quantity of the undecomposed system.

Concerning the power exchanges, additive or not depends on the different work variables \R{38d}
which generate external and internal power exchanges, \R{38d4} and \R{38d5}. Concerning the heat
exchanges, the split into external and internal ones is achieved by that of the propagator \R{+49}.
External power and heat exchanges are additive according to \R{38d6}$_1$ and \R{1dL3ax}$_1$,
whereas internal power and heat exchanges, \R{38e1} and \R{2dL3ax}, are not additive, caused by
the partition between the sub-systems.

Different contact temperatures of the sub-systems and the partition
between them prevent the additivity of the entropy exchanges generating the inequalities
\R{EX2} and \R{EX§1}. 
In general, the entropy \R{+117} and the entropy time rates \R{+131} are not additive unless the
density ope\-rator of the undecomposed system decomposes into those of the sub-systems \R{131a}.

More general:
If $\boxtimes$ is a quantity of the undecomposed system and $\boxtimes^1$ and $\boxtimes^2$
are the corresponding quantities of the sub-systems after the decomposition of the undecompossed
system into a bipartite one, then the {\em compound deficiency} $\boxtimes_{cd}$ is defined as follows \C{MUBE04}
\bee{68n}
\boxtimes_{cd}\ :=\ \boxtimes - \boxtimes^1 - \boxtimes^2\ =:\
\boxtimes - \boxtimes^{da}.
\ee
Here, $\boxtimes^{da}$ is the quantity which is generated by decomposed additivity.

Consequently, there are two possibilities to describe a bipartite system: as a decomposed one by
$\boxtimes^{da}$ or as an undecomposed one by $\boxtimes$. These descriptions are of different
information about the system, if the compound deficiency $\boxtimes_{cd}$ is not zero.
The answer to the question "What is the correct entropy time rate of the system?" depends
according to \R{+131} on its description chosen as decomposed or undecomposed.

According to \R{68n}, we obtain from \R{§1} the compound deficiency of the Hamiltonian
\bee{68o}
{\cal H}_{cd}\ :=\ {\cal H} - {\cal H}^1 - {\cal H}^2\ =\ {\cal H}^{12},
\ee
which is caused by the interaction part, a fact which is obvious: the sum of the
Hamiltonians of the sub-sytems differs from the Hamiltonian of the corresponding
undecomposed system.

\subsection{Removing semi-classical description\label{RCD}}

Up to now, a decomposed bipartite system is described semi-classically, that means, the external
exchange quantities with its environment are not connected to an interaction Hamiltonian, but are
represented by the exchange propagator \R{+48}. Now, this environment is separated from
the decomposed system by isolation
\bee{L7}
\ro_{ex}\ \equiv\ 0\ \longrightarrow\ \ro\ \equiv\ \ro_{iso},
\ee
thus considering an isolated bipartite system consisting of two irreversibly running 
sub-systems which are connected by an interaction Hamiltonian \R{68o}$_2$ which desribes 
exchange properties due to the partition between the two sub-systems. Consequently, the two
sub-systems are interacting, and the environment is isolated from them. The propagator $\ro$
belongs to the bipartite system and is as well as $\varrho$ decomposable into those of the interacting
sub-systems, \R{§2} and \R{§2a},
\bee{L8}
\ro{^1}\ :=\ \mbox{Tr}^2\ro,\qquad\ro{^2}\ :=\ \mbox{Tr}^1\ro.
\ee

Taking into account that the work variables are affected by the isolation according to \R{38d2}
\bee{L10}
\st{\td}{\mvec{a}}\!{^{1}_{iso}}\ \equiv\ \mvec{0},\qquad
\st{\td}{\mvec{a}}\!{^{2}_{iso}}\ \equiv\ \mvec{0},
\ee
and that the density operator $\varrho$ of the bipartite system is not affected by the isolation,
the in sect.\ref{DS} derived relations remain valid in the special case of isolation which is
characterized by \R{L7} and \R{L10}.

The exchanges between the sub-systems are by taking the partition \#(12) into account as follows:
\begin{itemize}
\item According to \R{38e1} and \R{38d5} the {\bf power exchange}
\bee{L11}
\Big\{
\mbox{Tr}^1\Big(\frac{{\partial\cal H}^1}{\partial\mvec{a}^{12}}\varrho^1\Big)
+\mbox{Tr}{^2}\Big(\frac{{\partial\cal H}^2}{\partial\mvec{a}^{12}}\varrho{^2}\Big)
+\mbox{Tr}\Big(\frac{{\partial\cal H}^{12}}{\partial\mvec{a}^{12}}
\varrho\Big)\Big\}
\cdot\st{\td}{\mvec{a}}\!{^{12}}\ =\ 0.
\ee
\item According to \R{MM2}, \R{MM4} and \R{2dL3ax}, the {\bf heat exchange}
\bey\nonumber
\mbox{Tr}\Big\{{\cal H}^1\Big(\ro-\frac{i}{\hbar}[{\cal H}^{12},\varrho]\Big)\Big\}
+\mbox{Tr}\Big\{{\cal H}^2\Big(\ro-\frac{i}{\hbar}[{\cal H}^{12},\varrho]\Big)\Big\}+
\\ \label{L12}
+\mbox{Tr}\Big\{{\cal H}^{12}\Big(\ro-\frac{i}{\hbar}
\Big[({\cal H}^{1}+{\cal H}^2),\varrho\Big]
\Big)\Big\}\ =\ 0.
\eey
\item According to \R{EX§1a}, \R{QU15y1} and \R{QU15y} the {\bf entropy exchange}
\bey\nonumber
\mbox{Tr}\Big\{\frac{{\cal H}^1}{\Theta^1}\Big(\ro-\frac{i}{\hbar}[{\cal H}^{12},\varrho]\Big)\Big\}
+\mbox{Tr}\Big\{\frac{{\cal H}^2}{\Theta^2}\Big(\ro-\frac{i}{\hbar}[{\cal H}^{12},\varrho]\Big)\Big\}\ \geq
\\ \label{L13}
\geq\ \frac{1}{T^1}\Omega^{int}(\frac{1}{\Theta^1}-\frac{1}{T^1})+\frac{1}{T^2}
\Omega^{int}(\frac{1}{\Theta^2}-\frac{1}{T^2})
\eey
which has different appearance due to the partition according to \R{EX§1}.
\item According to \R{=1}, \R{38d5}$_1$, \R{MM2}, \R{=2}$_1$, \R{38d5}$_2$, \R{MM4} and
\R{=3} the\newline {\bf energy exchange}
\byy{L14}
\st{\td}{E}{^A} = \mbox{Tr}^A\Big(\frac{{\partial\cal H}^A}{\partial\mvec{a}^{12}}\varrho^A\Big)\cdot\st{\td}{\mvec{a}}\!{^{12}}+
\mbox{Tr}\Big\{{\cal H}^A\Big(\ro-\frac{i}{\hbar}[{\cal H}^{12},\varrho]\Big)\Big\},\
A=1,2,\hspace{.3cm}
\\ \label{L15}
\st{\td}{E}{^{12}} = \mbox{Tr}\Big(\frac{{\partial\cal H}^{12}}{\partial\mvec{a}^{12}}
\varrho\Big)\cdot\st{\td}{\mvec{a}}\!{^{12}}+
\mbox{Tr}\Big\{{\cal H}^{12}\Big(\ro-\frac{i}{\hbar}
\Big[({\cal H}^{1}+{\cal H}^2),\varrho\Big]\Big)\Big\},\hspace{1cm}
\\ \label{L16}
\st{\td}{E}{^{1}}+\st{\td}{E}{^{2}}+\st{\td}{E}{^{12}}\ =\ 
\mbox{Tr}\Big\{\frac{\partial}{\partial\mvec{a}^{12}}
\Big({\cal H}^1+{\cal H}^2+{\cal H}^{12}\Big)\varrho\Big\}\cdot\st{\td}{\mvec{a}}\!{^{12}}\ =\ 0.
\hspace{.9cm}
\eey
\item According to \R{L5B} and \R{L5c} the {\bf entropy rate} and {\bf -production}
\byy{L16a}
\st{\td}{S}_A &=&  -k_B\mbox{Tr}\Big\{\Big(\ro-\frac{i}{\hbar}\Big[{\cal H}^{12},\varrho\Big]\Big)\ln(Z\varrho^A)\Big\},\quad A=1,2,\ C\neq A,\hspace{.8cm}
\\ \label{L16b}
\Sigma_ A &=&
-\mbox{Tr}\Big\{\Big(\frac{{\cal H}^A}{\Theta^A}+k_B\ln(Z\varrho^A)\Big) \Big(\ro
-\frac{i}{\hbar} \Big[{\cal H}^{12},\varrho\Big]\Big)\Big\}=
\\ \label{L16b1}
&=&
-\mbox{Tr}^A\Big\{\Big(\frac{{\cal H}^A}{\Theta^A}+k_B\ln(Z\varrho^A)\Big)\mbox{Tr}^C
\Big(\ro-\frac{i}{\hbar} \Big[{\cal H}^{12},\varrho\Big]\Big)\Big\}.
\eey

\end{itemize}

Although the contact temperature cannot be defined in original quantum mechanics
($\ro\equiv0$),  the thermodynamical quantities entropy rate and -production, heat- and
entropy exchange are established for bipartite systems by ${\cal H}^{12}$ also in this case.
The contact temperature which appears according to \R{L13} and \R{L16b} in entropy exchange
and entropy production is for original quantum mechanics added by hand from phenomenological
non-equilibrium thermodynamics. No such problems arise, if quantum thermodynamics takes the propagator $\ro$ into consideration modifying the von Neuman equation.

According to \R{L7} and \R{L10}, the relations \R{L11} to \R{L16b1} belong to an isolated bipartite
system whose sub-systems, \#1 and \#2, are interacting with each other through a material
impervious partition described by the interaction Hamiltonian ${\cal H}^{12}$. We now interprete
these sub-systems for getting rid of the semi-classical description of the bipartite system:
\byy{L17}
\#1\ &\longrightarrow&\ \mbox{the system}\ {\cal G},
\\ \label{L18}
\#2\  &\longrightarrow&\ \mbox{the environment}\ {\cal G}^\Box,
\\ \label{L19}
\#(12)\ &\longrightarrow&\ \mbox{the partition}\ \partial {\cal G},
\\ \label{L20}
\varrho,\ \ro\  &\longrightarrow&\ \mbox{belonging to the undecomposed sytem}\ 
{\cal G}\cup{\cal G}^\Box,
\\ \label{L20a}
\varrho{^A},\ \ro{^A}\  &\longrightarrow&\ \mbox{belonging to the sub-system \#A}.
\eey

With that, the semi-classical description of sub-systems in contact is replaced by a 
quantum-thermal one represented by \R{L11} to \R{L16b1}. Special contacts, such as usually
considered in quantum thermodynamics \C{KOS13}, are considered in the next section.

\subsubsection{Inert contacts\label{IC}}

Inert partitions between sub-systems establish inert contacts which are defined by equal
incoming and outgoing power- and heat exchanges through $\partial{\cal G}$, according to
\R{L11} and \R{L12} [and \R{MMe}] resulting in
\bee{L21}
\mbox{Tr}\Big(\frac{{\partial\cal H}^{12}}{\partial\mvec{a}^{12}}
\varrho\Big)\cdot\st{\td}{\mvec{a}}\!{^{12}}\ =\ 0,\quad
\mbox{Tr}\Big\{{\cal H}^{12}\Big(\ro-\frac{i}{\hbar}
\Big[({\cal H}^{1}+{\cal H}^2),\varrho\Big]\Big)\Big\}\ =\ 0.
\ee
Consequently, the two first terms in \R{L11} and \R{L12} remain, and $\st{\td}{E}\!{^{12}}=0$
is zero.

Usually, in quantum thermodynamics not only inert but also mono-sheet partitions are considered.
Consequently, \R{QU12c} and \R{QU12d} are valid, and \R{L13} yields
\bee{L22}
\mbox{Tr}\Big\{\frac{{\cal H}^1}{\Theta^1}\Big(\ro-\frac{i}{\hbar}[{\cal H}^{12},\varrho]\Big)\Big\}
+\mbox{Tr}\Big\{\frac{{\cal H}^2}{\Theta^2}\Big(\ro-\frac{i}{\hbar}[{\cal H}^{12},\varrho]\Big)\Big\}\ \geq\ 0,
\ee
that means, because of different contact temperatures of the sub-systems,
the entropy exchange is discontinuous at $\partial{\cal G}$. Taking \R{L12} and
\R{L21}$_2$ into account, \R{L22} results in
\bee{L23}
\mbox{Tr}\Big\{\Big(\frac{1}{\Theta^2}-\frac{1}{\Theta^1}\Big){\cal H}^2
\Big(\ro-\frac{i}{\hbar}[{\cal H}^{12},\varrho]\Big)\Big\}\ \geq\ 0,
\ee
or using \R{AL3a}$_4$
\bee{L24}
\Big(\frac{1}{\Theta^2}-\frac{1}{\Theta^1}\Big)\st{\td}{Q}\!{^2}\ \geq\ 0,
\ee
an inequality which is analogous to the defining inequalities \R{DU3c} and \R{NJ1}.

\subsubsection{Heat reservoirs\label{HRDI}}

We now choose the environment \#2 as a heat reservoir (HR) which is cha\-racterized by the
following properties: It is a quasi-static system of the thermostatic temperature $T_{HR}$
whose density operator is form-invariantly canonical during the contact time between \#1 and HR
according to \R{52d} 
\byy{L24a}
\varrho{^2_{HR}} &=& \frac{1}{Z} 
\exp\Big[-\frac{{\cal H}{^2_{HR}}}{k_B T_{HR}}\Big],
\\ \label{L24b}
Z &=& \mbox{Tr}^2\exp\Big[-\frac{{\cal H}{^2_{HR}}}{k_B T_{HR}}\Big],\quad
[{\cal H}{^2_{HR}},\varrho{^2_{HR}}]=\underline{0}.
\eey
Quasi-static means that non-equilibrium variables are not taken into account, and $T_{HR}$ is the
only slowly time dependent equilibrium variable according to \R{L24a}.
The heat reservoir undergoes a slow reversible (quasi-static) process \C{MU21}, whereas the
state of the system \#1 may change irreversibly. 

Contacting the heat reservoir \#2 with the the system \#1 through an inert and mono-sheet
partition according to \R{EX§5}, exchanges (relative to the HR) happen according to \R{L11},
\R{L21}$_1$ and \R{L12}, \R{L21}$_2$ and \R{L12}
\byy{L25b}
\mbox{power exchange:}&\quad&\st{\td}{W}{^2_{HR}}\ =\ 
\mbox{Tr}{^2}\Big(\frac{{\partial\cal H}{^2_{HR}}}{\partial\mvec{a}^{12}}\varrho{^2_{HR}}\Big)\cdot\st{\td}{\mvec{a}}\!{^{12}},
\\ \label{L25c}
\mbox{heat exchange:}&\quad&\st{\td}{Q}{^2_{HR}}\ =\
\mbox{Tr}^2\Big\{{\cal H}{^2_{HR}}\Big(\ro\!{^2_{HR}}
-\frac{i}{\hbar}\mbox{Tr}^1\Big[{\cal H}^{12},\varrho\Big]\Big)\Big\},\hspace{.8cm}
\\ \label{L25d}
\mbox{entropy exchange:}&\quad&\Xi{^2_{HR}}\ =\ \st{\td}{Q}{^2_{HR}}/T_{HR}.
\eey
The heat exchange between HR and \#1 is according to \R{L25c} determined by the propagator
$\ro\!\!{^2_{HR}}$ which appears in the corresponding modified von Neumann equation \R{+45}.
By taking \R{L24b}$_2$ into account, \R{+45} results in
\bee{L25}
\st{\td}{\varrho}{^2_{HR}}\ =\ -\frac{i}{\hbar}\mbox{Tr}^1 \Big[{\cal H}^{12},\varrho\Big]+\ro{^2_{HR}},
\ee
whereas $\st{\td}{\varrho}\!{^1}$ is determined by \R{+44}.

Because of the presupposed form-invariance of the canonical density ope\-rator \R{L24a} during the
thermal contact between HR and \#1, an other expression for $\st{\td}{\varrho}\!{^2_{HR}}$
can be generated by differentiation of \R{L24a} which is done in sect.\ref{HR}. The result is
\bee{zL25}
\st{\td}{\varrho}\!{^2_{HR}}\ =\ 
\varrho{^2_{HR}}\Big\{\mbox{Tr}^2\Big(\varrho{^2_{HR}}\ln(Z\varrho{^2_{HR}})\Big)
-\ln(Z\varrho{^2_{HR}})\Big\}\frac{\st{\td}{T}_{HR}}{T_{HR}}\ =:\
{\cal C}{^2_{HR}}\st{\td}{T}_{HR}.
\ee
Comparing \R{zL25} with \R{L25} results in the propagator of the HR
\bee{tL25}
\ro\!{^2_{HR}}\ =\ \frac{i}{\hbar}\mbox{Tr}^1 \Big[{\cal H}^{12},\varrho\Big]+
{\cal C}{^2_{HR}}\st{\td}{T}_{HR},
\ee
and the heat exchange \R{L25c} between HR and \#1 becomes
\bee{sL25}
\st{\td}{Q}{^2_{HR}}\ =\
\mbox{Tr}^2\Big({\cal H}{^2_{HR}}{\cal C}{^2_{HR}}\Big)\st{\td}{T}_{HR}\
=:\ C_{HR}\st{\td}{T}_{HR}.
\ee
The reversible process of the HR is caused by its huge heat capacity\newline 
${C_{HR}{-\!\!\!>}\infty}$ with respect to that of the contacted system \#1
resulting in ${\st{\td}{T}\!\!{_{HR}}{-\!\!\!>}0}$,
a property which characterize the HR as an idealized system\footnote{$C_{HR}>0$ is not proven}.

Because the considered partition between HR and \#1 is presupposed as being inert,
$-\st{\td}{Q}\!{^1}=\ \st{\td}{Q}{^2_{HR}}$,
\R{L21}$_2$ and \R{L12} result with \R{L25c} and \R{sL25}$_2$ in
\bey\nonumber
-\mbox{Tr}^1\Big\{{\cal H}{^1}\Big(\ro{^1}
-\frac{i}{\hbar}\mbox{Tr}^2\Big[{\cal H}^{12},\varrho\Big]\Big)\Big\}\ =\ 
C_{HR}\st{\td}{T}_{HR}\ =\ 
\\ \label{L25e}
=\ \mbox{Tr}^2\Big\{{\cal H}{^2_{HR}}\Big(\ro{^2_{HR}}
-\frac{i}{\hbar}\mbox{Tr}^1\Big[{\cal H}^{12},\varrho\Big]\Big)\Big\}.\hspace{-1cm}
\eey
As expected, contacting HR with \#1 and $\st{\td}{T}_{HR}$ is small positive, then $T_{HR}$ is
smaller than the contact temperature $\Theta^1$ of \#1, thus \R{L24} is satisfied
\bee{L27}
\Big(\frac{1}{T_{HR}}-\frac{1}{\Theta^1}\Big)\st{\td}{Q}\!{^2_{HR}}\ \geq\ 0,\qquad
\Big(\frac{1}{\Theta^1}-\frac{1}{T_{HR}}\Big)\st{\td}{Q}\!{^1_{HR}}\ \geq\ 0.
\ee

\subsubsection{Equilibrium}

Equilibrium is defined by equilibrium conditions which are divided into ne\-cessary and complementary
ones. The necessary equilibrium conditions for {\bf undecomposed systems} are time independent
work variables and vanishing entropy production
\bee{L31b}
\st{\td}{\mvec{a}}_{eq}\ \equiv\ \mvec{0}\ \longrightarrow\ \st{\td}{W}_{eq}\ =\ 0,\quad 
\Sigma_{eq}\ \equiv\ 0\ \longrightarrow\ 0\ =\ \Xi_{eq}-\st{\td}{S}_{eq}.
\ee
We obtain from \R{+18} and \R{+15} the necessary equilibrium condition \R{L31b}$_4$
\bee{L31c}
0\ =\ 
\mbox{Tr}\Big\{\Big(\frac{{\cal H}_{eq}}{\Theta_{eq}}+k_B\ln(Z\varrho_{eq})\Big)\ro_{eq}\Big\}.
\ee
The left hand bracket in \R{L31c} set to zero, results by use of \R{+6}$_1$ in a first complementary
equilibrium condition (\R{L31b}$_1$ is silent valid)
\bee{L31d}
\varrho_{eq}\ =\ \frac{1}{Z}\exp\Big(-\frac{{\cal H}_{eq}}{k_B\Theta_{eq}}\Big)\
\longrightarrow\ [{\cal H}_{eq},\varrho_{eq}]\ =\ \underline{0}\ \longrightarrow\
\st{\td}{\varrho}_{eq}\ =\ \ro_{eq}
\ee
which is not sufficient because of \R{L31d}$_3$. Adding a second complementary equilibrium
condition, $\ro_{eq}\doteq\underline{0}$, (which alone is also not sufficient) makes both of them
\bee{L31d3}
\frac{{\cal H}_{eq}}{\Theta_{eq}}+k_B\ln(Z\varrho_{eq})\ =\ 0
\quad \wedge\quad\ro_{eq}\ =\ 0
\ee
sufficient for equilibrium because they induce according to \R{+6}
to \R{+18}
\bee{L31d4}
\st{\td}{\varrho}_{eq}\ =\ 0,\quad\st{\td}{E}_{eq}\ =\ 0,\quad \st{\td}{Q}_{eq}\ =\ 0,\quad
\st{\td}{S}_{eq}\ =\ 0,\quad\Xi_{eq}\ =\ 0,\quad\Theta_{eq}\ =\ T^\Box.
\ee
Consequently, the necessary equilibrium conditions \R{L31b}, especially the vani\-shing entropy
production are not sufficient for equilibrium. Sufficient for equilibrium are the joint supplementary
equilibrium conditions \R{L31d3}.

Equilibrium of {\bf bipartite systems} means endoreversible equilibrium: sub-systems are in
equilibrium and exchanges between them are zero. The necessary equilibrium conditions are
according to \R{L11}, \R{L16b1} and \R{+20}
\byy{L31d5}
\st{\td}{\mvec{a}}{^{12}_{eq}} &\equiv& \mvec{0}\ \longrightarrow\
\st{\td}{W}{^A_{eq}}\ =\ 0,
\\ \label{L31d6}
\Sigma{_A^{eq}} &=& -\mbox{Tr}^A\Big\{\Big(\frac{{\cal H}{^A_{eq}}}{\Theta{^A_{eq}}}+k_B\ln(Z\varrho{^A_{eq}})\Big)\mbox{Tr}^C
\Big(\ro_{eq}-\frac{i}{\hbar} \Big[{\cal H}{^{12}_{eq}},\varrho_{eq}\Big]\Big)\Big\}\equiv\ 0,
\hspace{.7cm}
\\ \label{aL31d6}
&&\longrightarrow\ 0\ =\ \Xi{^A_{eq}}-\st{\td}{S}{^A_{eq}},\qquad\qquad A=1,2,\ C\neq A,
\eey
resulting from \R{L31d6} in two supplementary equilibrium conditions for vanishing entropy
production according to \R{+44}, \R{+45} and \R{+6}$_1$
\byy{L31d7}
\ro{^A_{eq}}-\frac{i}{\hbar}\mbox{Tr}^C \Big[{\cal H}{^{12}_{eq}},\varrho_{eq}\Big]\ =\
\underline{0},\ 
\longrightarrow\ \st{\td}{\varrho}{_{eq}^A}\ =\ 
-\frac{i}{\hbar} [{\cal H}{_{eq}^A},\varrho{_{eq}^A}],
\\ \nonumber
\varrho{_{eq}^A}\ =\ \frac{1}{Z}\exp\Big(-\frac{{\cal H}{_{eq}^A}}{k_B\Theta{_{eq}^A}}\Big)\
\longrightarrow\ [{\cal H}{_{eq}^A},\varrho{_{eq}^A}]\ =\ \underline{0}\ \longrightarrow
\\ \label{L31d8}
\longrightarrow\ \st{\td}{\varrho}{_{eq}^A}\ =\ 
-\frac{i}{\hbar}\mbox{Tr}^C  \Big[{\cal H}{^{12}_{eq}},\varrho_{eq}\Big]+\ro{_{eq}^A}.
\eey
The interaction of the two supplimentary equilibrium conditions \R{L31d7} and \R{L31d8} is
remarkable: Although each of them gives rise to vanishing entropy production, equilibrium only
exists, if both are valid
\bee{aL31d8}
\st{\td}{\varrho}{_{eq}^A}\ =\ \underline{0}\ =\ 
-\mbox{Tr}^C \Big[{\cal H}{^{12}_{eq}},\varrho_{eq}\Big]+\ro{_{eq}^A},
\qquad A=1,2,\ C\neq A,
\ee
and if additionally the partial propagator vanishes
\bee{aL31d8a}
\ro{_{eq}^A}\ =\ \underline{0}\ \longrightarrow\ 
\mbox{Tr}^C \Big[{\cal H}{^{12}_{eq}},\varrho_{eq}\Big]\ =\ \underline{0},
\ee
resulting in a constraint for the interaction Hamiltonian in equilibrium. Clear is that vanishing entropy
production is not sufficient for equilibrium. As \R{L31d6} and \R{aL31d6}, also \R{aL31d8} and
\R{aL31d8a} are a necessary equilibrium conditions. A further one is according to \R{aL31d6}
\bee{bL31d8}
\st{\td}{S}{^A_{eq}}\ \doteq\ 0\ \longrightarrow\ \Xi{^A_{eq}}\ =\ 0\ \longleftrightarrow\ 
\st{\td}{Q}{^A_{eq}}\ =\ 0.
\ee
The set of all necessary equilibrium conditions is sufficient for equilibrium.

Because the heat exchange is a continuous and monotonous function according to \R{DU10c}, from
\R{L24} and \R{bL31d8}$_3$ follows the thermostatic temperature $T$ in equilibrium
\bee{L31d9}
\Theta{^1_{eq}}\ =\ \Theta{^2_{eq}}\ =:\ T.
\ee

The in \ref{IC} presupposed conditions for inert contacts \R{L21} are satisfied also in equilibrium what can be confirmed by
inserting \R{L31d7}$_1$ into the first two terms of \R{L12} resulting in \R{L21}$_2$.

 \section{Summary and Survey}

Deriving the von Neumann equation from the Schr\"odinger equation in ori\-ginal quantum
mechanics, the time rates of the weights of the statistical operator are set to zero without any 
argumentation. If these time rates are taken into account, a modified von Neumann equation
is generated which allows to introduce thermodynamical properties into quantum mechanics,
as demonstrated in the first part of this paper \C{MU19,MU20} for undecomposed Schott\-ky systems.
The second part deals with closed decomposed Schott\-ky systems, especially with bipartite systems
interacting with an environment semi-classically and with isolated bipartite systems with interacting
sub-systems. After a brief repetition of thermodynamics of undecomposed systems, that of
closed decomposed systems is discussed in the following steps
\begin{itemize}
\item
The {\em von Neumann equation} is modified by a so-called propagator $\ro$ including the time
rates of the density operator's weights which are set to zero in original quantum mechanics. Density
operator and propagator are adapted to the considered bipartite system by tracing, thus generating
the equations of motion of the traced density operators belonging to the two sub-systems of the
bipartite system (sect.\ref{MVNE}).
\item
The {\em exchanges} between the bipartite system and its environmement and between the two
sub-systems are taken into consideration (sect.\ref{EXCH}). There are three kinds of exchanges for
closed Schottky systems:$\!$ power-, heat- and entropy-exchange which are defined by the two
partial Hamiltonians of the bipartite system and the interaction Hamiltonian belonging to the
sub-systems:
${\cal H}\ =\ {\cal H}^1+{\cal H}^2+{\cal H}^{12}$.
\begin{itemize}
\item[$\Box$]
For describing the {\em power exchange}, two kinds of work variables are introduced
(sect.\ref{PEX}):
those which belong to the external power exchange between the decomposed system and its
environment ($\st{\td}{\mvec{a}}\!{^1}$ and $\st{\td}{\mvec{a}}\!{^2}$)
and those which describe the internal power exchange between the sub-systems of the bipartite
system ($\st{\td}{\mvec{a}}\!{^{12}}$). Presupposing the dependence of the partial Hamiltonians
on the work variables
\bey\nonumber
{\cal H}^1(\mvec{a}^1,\mvec{a}^{12}),\quad {\cal H}^2(\mvec{a}^2,\mvec{a}^{12}),\quad {\cal H}^{12}(\mvec{a}^{12}),
\eey
the external power exchange is \R{38d4}
\bey\nonumber
\st{\td}{W}_{ex}\ =\ 
\sum_A\mbox{Tr}^A\Big(\frac{\partial {\cal H}^A}{\partial\mvec{a}^A}\varrho^A\Big)\cdot
\st{\td}{\mvec{a}}\!{^A},\qquad A=1,2,
\eey 
and the internal power exchange \R{38d5}, \R{38d6}$_2$ and \R{38e} vanishes
\bey\nonumber
\st{\td}{W}_{int}\ =\ 
\sum_B\mbox{Tr}\Big(\frac{\partial {\cal H}^B}{\partial\mvec{a}^{12}}\varrho\Big)\cdot
\st{\td}{\mvec{a}}\!{^{12}}\ \equiv\ 0,\qquad B=1,2,(12).
\eey
\item[$\Box$]
Starting with the definition of {\em heat exchange} \R{AL3a} and with the modified von Neumann equation
\R{+43}
\bey\nonumber
\st{\td}{Q}{^B}\ :=\ \mbox{Tr}({\cal H}^B\st{\td}{\varrho})\ =\ 
\mbox{Tr}\Big\{{\cal H}^B \Big(\ro
-\frac{i}{\hbar} \Big[{\cal H},\varrho\Big]\Big)\Big\},\qquad B=1,2,(12),
\eey
the sum of the heat exchanges results in
\bey\nonumber
\sum_B\st{\td}{Q}{^B}\ =\ \mbox{Tr}({\cal H}\ro).
\eey
The heat exchange has to be split into its external and its internal part.
Whereas the splitting into external and internal exchange is achieved for the power exchange by
the work variables, for the heat exchange this splitting is done by that of the propagator \R{+49} 
\bey\nonumber
\ro\ = \ \ro_{ex} + \ro_{iso},
\eey
at which $\ro_{iso}$ is generated by isolating the system. Because ${\cal H}{^{12}}$ belong to the
partition separating the two sub-systems which has no contact to the system's environment
\R{aL3a3}$_2$
\bey\nonumber
\mbox{Tr}({\cal H}{^{12}}\ro_{ex}) \equiv\ 0\ \longrightarrow\ 
\st{\td}{Q}{^{1}_{ex}}+\st{\td}{Q}{^{2}_{ex}}\ =\ \st{\td}{Q}_{ex},
\eey
and because the sum of all internal heat exchanges is zero \R{aL3a3}$_1$
\bey\nonumber
\mbox{Tr}({\cal H}\ro_{iso})\equiv 0\ 
\longrightarrow\ 
-\st{\td}{Q}{^{1}_{int}}\ =\ \st{\td}{Q}{^{2}_{int}}+\st{\td}{Q}{^{12}_{int}},
\eey
the external heat exchanges are additive, and the partition between the two sub-systems is in
general not inert ($-\st{\td}{Q}{^{1}_{int}}\neq\st{\td}{Q}{^{2}_{int}}$). 
\item[$\Box$]
In the realm of discrete systems, the {\em entropy exchange} is defined as heat exchange over 
temperature \{\R{IM8}$_1$ and \R{EX§1a}$_3$ (sect.\ref{EEX})\footnote{This may be different in
field formulation of thermodynamics because more exchanges appear as those for Schottky
systems.}\}. The contact temperature is defined by the defining inequalities \R{DU3a},\R{NJ1}
\C{MU09,MU18a}. Four temperatures appear in bipartite systems: $\Theta^1,\Theta^2$ and
$\Theta^{12}$, the contact temperatures of the sub-systems and that of the partition between
them and $T^\Box$ or $T_{HR}$ the temperature of the system's environment. Because of these
different temperatures, the entropy exchanges are not additive \R{EX2} and not continuous even
at inert partitions \R{L22}\footnote{for which power- and heat-exchange are continuous}.
\end{itemize}
\item
{\em Partial density operators} are defined by tracing the density operator of the undecomposed system,
\R{§2} and \R{§2a}. Consequently, the definition of the Shannon entropy of the undecomposed
system can be transfered to the sub-systems, with the result that the sum of the partial entropies
is not smaller than that of the undecomposed system. Also the sum of the partial entropy time
rates is different from the entropy time rate of the undecomposed system\footnote{an effect called:
compound deficiency}.
\item
{\em Partial entropy productions} are defined by the partial entropy time rates  minus the
partial entropy exchanges \R{L5a}. The quantum-thermal expression of the {\em partial contact
temperature} of a sub-system \R{c?1ay} is achieved by three procedures according to \R{L5c2}:
separation of the sub-systems ${\cal H}^{12}\doteq \underline{0}$, replacing
$\ro\!{^A}\rightarrow\ \ro\!{^A}\!-\varrho{^A_{iso}}=\varrho{^A_{ex}}$
and setting its partial entropy production $\Sigma{^0_A}$ to zero which results in \R{L5b}.
The entropy production of separated sub-systems vanishes in original quantum mechanics according
to \R{L5c2}, whereas in quantum thermodynamics it does not. 
\item
Sufficient {\em equilibrium conditions} for undecoposed systems are according to \R{L31d3}
the vanishing propagator and the canonical statistical operator. For the sub-system \#A of a bipartite
system follows from the vanishing entropy production two necessary equilibrium conditions, the
equation of "motion" of the statistical operator of reversible "processes" \R{L31d7}$_2$ and
the canonical shape of this statistical operator \R{L31d8}$_1$. Because the entropy time rate
vanishes in equilibrium, heat exchange and entropy exchange vanish, too, representing additional
necessary equilibrium conditions which all together are sufficient for equilibrium.
\item
A {\em heat reservoir} is defined as a quasi-static system of the thermostatic temperature $T_{HR}$
whose density operator is form-invariantly canonical during the contact time between the system
and the heat reservoir \R{L24a}. This definition results in a special time rate of the statistical
operator of the heat reservoir \R{zL25} which is compared with that of a sub-system \R{tL25}.  
Additionally is presupposed that the heat capacity of the heat reservoir is huge compared with that
of the contacted system, resulting in a tiny time rate of the reservoir's thermostatic temperature.
Presupposing an inert partition between heat reservoir and system, the heat exchange between
them satisfies a defining inequality \R{L27}.

\end{itemize}

\section{Appendices}
\subsection{Tracing\label{TRA}}

Starting with a tensor of the decomposed system
\bee{3.1}
A\ =\
\sum_{kl}\sum_{pq}|\Psi_1^k>|\Psi_2^l>A^{pq}_{kl}<\Psi_2^q|
<\Psi_1^p|,
\ee
and its traces
\byy{3.2}
\mbox{Tr}^1 A\ =\ \sum_{lq} |\Psi_2^l> \Big(\sum_m
A^{mq}_{ml}\Big)<\Psi_2^q|\ =:\ A^2,
\\ \label{3.2a}
\mbox{Tr}^2 A\ =\ \sum_{kp} |\Psi_1^k> \Big(\sum_m
A^{pm}_{km}\Big)<\Psi_1^p|\ =:\ A^1.
\eey
Considering the density operator \R{38a} and the propagator \R{38b}, we obtain two special cases
\byy{a3.2a}
A\ \doteq\ \varrho,\quad q\ \doteq\ l,\quad p\ \doteq\ k,\quad A^{pq}_{kl}\ =\  A^{kl}_{kl}\ =\ 
p_{kl},\quad\sum_{jm} p_{jm}\ =\ 1,
\\ \label{b3.2a}
A\ \doteq\ \ro,\quad q\ \doteq\ l,\quad p\ \doteq\ k,\quad A^{pq}_{kl}\ =\  A^{kl}_{kl}\ =\ 
\st{\td}{p}_{kl},\quad\sum_{jm} \st{\td}{p}_{jm}\ =\ 0,
\\ \label{c3.2a}
\varrho^1 = \sum_k |\Psi_1^k>\!\Big(\sum_jp_{kj}\Big)\!<\Psi_1^k|,\ \ 
\varrho^2 = \sum_l |\Psi_2^l>\!\Big(\sum_jp_{jl}\Big)\!<\Psi_2^l|,
\\ \label{d3.2a}
\ro\!{^1} = \sum_k |\Psi_1^k>\!\Big(\sum_j\st{\td}{p}_{kj}\Big)\!<\Psi_1^k|,\ \ 
\ro\!{^2} = \sum_l |\Psi_2^l>\!\Big(\sum_j\st{\td}{p}_{jl}\Big)\!<\Psi_2^l|,
\eey

Because partial traces commute, we obtain
\byy{59y}
\mbox{Tr}^2 A^2\ =\ \mbox{Tr}^2\mbox{Tr}^1 A \ =\ \mbox{Tr}A\ =\ 
\mbox{Tr}^1\mbox{Tr}^2 A\ =\ \mbox{Tr}^1 A^1\ =\ \sum_{mr}A^{rm}_{rm},
\\ \label{a61}
A\ \doteq\ \varrho\longrightarrow\mbox{Tr}^2\!\ro{^2}\ =\ \mbox{Tr}\varrho\ 
=\ 1\ =\ \mbox{Tr}^1\!\ro{^1},
\\ \label{3.3}
\mbox{Tr}(A^1 B)\ =\ \mbox{Tr}^1\mbox{Tr}^2(A^1B)\ =\
\mbox{Tr}^1(A^1B^1).
\eey
A detailled calculation
\bey\nonumber
\mbox{Tr}^1(A^1B)\ =\
\sum_j\sum_{sr}\sum_{kl}\sum_{pq}\hspace{6cm}
\\ \nonumber
<\Psi{^j_1}|\Psi{^s_1}>A{^r_s}
<\Psi{^r_1}|\Psi{^k_1>}|\Psi{^l_2>}B{^{pq}_{kl}}<\Psi{^q_2}|
<\Psi{^p_1}|\Psi{^j_1}>\ =\ 
\\ \label{59z}
=\ \sum_j\sum_k\sum_{ql}   A{_j^k} B{^{jq}_{kl}}|\Psi{^l_2>}<\Psi{^q_2}|,
\eey
\bey\nonumber
\mbox{Tr}^1(BA^1)\ =\
\sum_j\sum_{kl}\sum_{pq}\sum_{sr}\hspace{6cm}
\\ \nonumber
<\Psi{^j_1}|\Psi{^k_1>}|\Psi{^l_2>}B{^{pq}_{kl}}<\Psi{^q_2}|<\Psi{^p_1}|
\Psi{^s_1}>A{^r_s}<\Psi{^r_1}|\Psi{^j_1}>\ =\ 
\\ \label{59r}
=\ \sum_j\sum_p\sum_{ql}B{^{pq}_{jl}}A{_p^j}|\Psi{^l_2>}<\Psi{^q_2}|\ =\ 
\sum_j\sum_k\sum_{ql}B{^{jq}_{kl}}A{_j^k}|\Psi{^l_2>}<\Psi{^q_2}|
\eey
results in
\bee{59s}
\mbox{Tr}^1[A^1,B]\ =\ 0\ \longrightarrow\ \mbox{Tr}^2\mbox{Tr}^1[A^1,B]\ =\ 
\mbox{Tr}^1[A^1,B^1]\ =\ 0.
\ee

\subsection{Separation Axiom\label{SA}}

Starting with \R{aL5} and \R{bL5}
\byy{=aL5}
\st{\td}{Q}\!{^A_{sep}}&\doteq& \mbox{Tr}^A({\cal H}^A\ro\!{^A_{sep}}),
\\ \label{=bL5}
\st{\td}{S}{_A^{sep}}&\doteq& -k_B\mbox{Tr}^A\{\ro\!{^A_{sep}}\ln(Z\varrho^A)\},
\eey
and inserting \R{cL5}
\bee{=cL5}
\ro\!{^A_{sep}}\ =\ \Big[{\cal H}^A,\Big[\ln(Z\varrho{^A}),{\cal H}^A\Big]\Big]\ =:
\Big[{\cal H}^A,X^A\Big]
\ee
into \R{=aL5} and \R{=bL5} results in
\byy{=dL5}
\st{\td}{Q}\!{^A_{sep}}&\doteq& \mbox{Tr}^A\Big({\cal H}^A\Big[{\cal H}^A,X^A\Big]\Big) =
\mbox{Tr}^A\Big({\cal H}^A{\cal H}^A X^A-{\cal H}^A{\cal H}^AX^A\Big) = 0,
\hspace{.4cm}
\\ \nonumber
\st{\td}{S}{_A^{sep}}&\doteq&
-k_B\mbox{Tr}^A\Big\{\Big[{\cal H}^A,X^A\Big]\ln(Z\varrho^A)\Big\}\ =\
\\ \label{=eL5}
&=&-k_B\mbox{Tr}^A\Big\{X^A\Big[\ln(Z\varrho^A),{\cal H}^A\Big]\Big\}.
\eey
Inserting the definition of $X^A$ \R{=cL5}, the entropy time rate results in
\bee{=dL5}
\st{\td}{S}{_A^{sep}}\ =\
-k_B\mbox{Tr}^A\Big\{\Big[\ln(Z\varrho{^A}),{\cal H}^A\Big]
\Big[\ln(Z\varrho{^A}),{\cal H}^A\Big]\Big\}\ \neq\ 0.
\ee

\subsection{Heat Reservoirs\label{HR}}

Because there is no power exchange
$\st{\td}{\mvec{a}}\!{^{12}}=\mvec{0}$ between the HR and the system \#1,
$\st{\td}{\cal H}{^2_{HR}}=\underline{0}$ is valid, and the 
differentiation of the canonical statistical operator \R{L24a}
\bee{*L24a}
\varrho{^2_{HR}}\ =\ \frac{1}{Z}\exp\Big[-\frac{{\cal H}{^2_{HR}}}{k_B T_{HR}}\Big]
\ee
results in
\bey\nonumber
\st{\td}{\varrho}\!{^2_{HR}} &=& -\frac{1}{Z^2}\st{\td}{Z}
\exp\Big[-\frac{{\cal H}{^2_{HR}}}{k_B T_{HR}}\Big]
+\frac{1}{Z}\exp\Big[-\frac{{\cal H}{^2_{HR}}}{k_B T_{HR}}\Big]
\Big(-\frac{{\cal H}{^2_{HR}}}{k_B}\Big(-\frac{\st{\td}{T}_{HR}}{T{^2_{HR}}}\Big)
\\ \label{*L24a+}
&=&-\frac{\st{\td}{Z}}{Z}\varrho{^2_{HR}}+\varrho{^2_{HR}}
\Big(\frac{{\cal H}{^2_{HR}}}{k_BT_{HR}}\frac{\st{\td}{T}_{HR}}{T_{HR}}\Big).
\eey
Following the differentiation of the normalization factor \R{L24b}
\bee{*L24b}
Z\ =\ \mbox{Tr}^2\exp\Big[-\frac{{\cal H}{^2_{HR}}}{k_B T_{HR}}\Big]
\ee
and taking the second term of \R{*L24b} into account, results in
\bee{*L24b+}
\st{\td}{Z}\ =\ \mbox{Tr}^2\Big\{Z\varrho{^2_{HR}}
\Big(\frac{{\cal H}{^2_{HR}}}{k_BT_{HR}}\frac{\st{\td}{T}_{HR}}{T_{HR}}\Big)\Big\}.
\ee
Inserting \R{*L24b+} into \R{*L24a+} and taking \R{*L24a} into account results in
\bey\nonumber
\st{\td}{\varrho}\!{^2_{HR}} &=& \varrho{^2_{HR}}\Big\{-\mbox{Tr}^2\Big(\varrho{^2_{HR}}
\frac{{\cal H}{^2_{HR}}}{k_BT_{HR}}\Big)+
\frac{{\cal H}{^2_{HR}}}{k_BT_{HR}}\Big\}
\frac{\st{\td}{T}_{HR}}{T_{HR}}\ =\
\\ \label{*L24b1+}
&=&
\varrho{^2_{HR}}\Big\{\mbox{Tr}^2\Big(\varrho{^2_{HR}}
\ln(Z\varrho{^2_{HR}})\Big)
-\ln(Z\varrho{^2_{HR}})\Big\}
\frac{\st{\td}{T}_{HR}}{T_{HR}}.
\eey

\end{document}